\documentclass[proceedings]{JHEP37}
\usepackage{eurosym}
\usepackage{amssymb}
\usepackage{amsfonts}
\usepackage{amsmath,epsfig}
\usepackage{graphicx}

\newbox\mybox

\newcommand\fverb{\setbox\mybox=\hbox\bgroup\verb}
\newcommand\fverbdo{\egroup\medskip\noindent\fbox{\unhbox\mybox}\ }
\newcommand\fverbit{\egroup\item[\fbox{\unhbox\mybox}]}
\conference{Nonlocal gauge equivalence}
\abstract{We exploit the gauge equivalence between the Hirota equation and the extended continuous Heisenberg equation to investigate how nonlocality properties of one system are inherited 
by the other. We provide closed generic expressions for nonlocal multi-soliton solutions for both systems. 
By demonstrating that a specific auto-gauge transformation for the extended continuous Heisenberg equation becomes
equivalent to a Darboux transformation, we use the latter to construct the nonlocal multi-soliton solutions
from which the corresponding nonlocal solutions to the Hirota equation can be computed directly. 
We discuss properties
and solutions of a nonlocal version of the nonlocal extended Landau-Lifschitz equation obtained from the nonlocal
extended continuous Heisenberg equation or directly from the nonlocal solutions of the Hirota equation.}

\title{Nonlocal gauge equivalence: Hirota versus extended continuous
Heisenberg and Landau-Lifschitz equation}
\author{Julia Cen$^\bullet$, Francisco Correa$^\circ$ and Andreas Fring$%
^\bullet$ \\
$\bullet$ Department of Mathematics, City, University of London,\\
$\,\,$ Northampton Square, London EC1V 0HB, UK \\
$\circ$ Instituto de Ciencias F{\'{\i}}sicas y Matem{\'{a}}ticas,
Universidad Austral de Chile, \\
$\,\,$ Casilla 567, Valdivia, Chile\\
E-mail: julia.cen.1@city.ac.uk, francisco.correa@uach.cl, a.fring@city.ac.uk}

\input{tcilatex}
\begin{document}

\section{Introduction}

Traditionally the study of solutions to nonlinear wave equations has mainly
focused on systems that involve local fields that all depend on a single
point in space-time. However, there are many well-known phenomena in nature
related to events that appear to be correlated to each other even though
they are separated in a spacelike or timelike fashion. In quantum mechanics
entanglement, see e.g. \cite{nielsen2010}, is a well studied phenomenon that
in some settings seems to exclude the possibility of signalling \cite%
{Bancal2016} and can be implemented even for many particle systems \cite%
{hosten2016}. There are, however, also physical phenomena that are of a
nonlocal nature that are describable with nonlinear wave equations, such as
nonlocal rogue waves \cite{rogue1,rogue2}, weather forecast models in which
nonlocality is caused by feedback loops \cite%
{jeffries2013arctic,overland2016melting}, gravitational waves \cite%
{abbott2016binary}, etc.

Recently a simple principle was identified \cite{ablowitz2013,ablowitz2016}
that introduces nonlocality into nonlinear integrable systems in a
systematic and mathematically well-defined manner. This is achieved by
exploring various versions of $\mathcal{PT}$-symmetry present in the zero
curvature condition that relates fields in the theory to each other in a
nonlocal fashion. One particular type of these new nonlocal nonlinear Schr%
\"{o}dinger equation has attracted a lot of attention \cite%
{nonlocalNLSE1,nonlocalNLSE2,nonlocalNLSE3,nonlocalNLSE4,nonlocalNLSE5,nonlocalNLSE6,nonlocalNLSE7,wen2016dynamics,gerdjikov2017complete}%
. These studies were extended to other types of systems, such as
Fordy-Kulish equations \cite{gurses2017nonlocal}, Davey-Stewartson equations 
\cite{rao2017rational,rao2018rogue}, Sasa-Satsuma equations \cite%
{song2017reverse}, Kadomtsev-Petviashvili equations \cite{zhang2017breather}
and Korteweg de-Vries systems \cite{AliceB1,AliceB2,AliceB3}. Here we will
build on a particular case of the various nonlocal versions of the Hirota
equation \cite{CenFringHir}. In the local case the extension from the
nonlinear Schr\"{o}dinger equation to the Hirota equation is suggested by
experiments in the high-intensity and short pulse subpicosecond regime \cite%
{mitschke1986,gordon1986} where the accurate description of the former
equation \cite{mollenauer1980} breaks down. The nonlocality is known to be
implementable by the applications of various variants of $\mathcal{PT}$%
-symmetry \cite{PTbook} to the equations resulting from an AKNS zero
curvature construction \cite{AKNS}. On the other hand one may also directly
decompose the fields in some local systems into field depending of different
points in space-time and obtain nonlocal systems in this manner, as
described in \cite{AliceB1,AliceB2,AliceB3}. Here we explore the possibility
to exploit the gauge equivalence of two systems and investigate how the
nonlocality properties of one system is inherited by the other. As concrete
systems we investigate the gauge equivalent pair of the extended versions of
the continuous limit of the Heisenberg equation (ECHE) \cite%
{nakamura1974solitons,lakshmanan1,tjon1977solitons,takhtajan197,demontis2018}
and the extended Landau-Lifschitz equations (ELLE) \cite%
{landau1935,bar2015landau}. The local version of the original
Landau-Lifschitz equation famously describes the precession of the
magnetization in a solid when subjected to a torque resulting from an
effective external magnetic field. Various extended versions have been
proposed, such as for instance the Landau-Lifshitz-Gilbert equation \cite%
{gilbert1955} to take damping into account. The nonlocal versions of these
equation studied here provide further extensions with complex components. We
will see how the nonlocality may be incorporated most naturally into a pair
of auxiliary equations occurring this setting.

We will also show how the nonlocality is implemented into two standard types
of solution procedures for nonlinear systems, Hirota's direct method and the
method of using Darboux transformations. Similarly as in \cite{CenFringHir}
we find new types of solutions in the nonlocal setting which have no
counterpart in the local case.

Our manuscript is organized as follows: In section 2 we introduce our three
systems the nonlocal Hirota equation, the extended versions of the
continuous limit of the Heisenberg equation and the extended Landau
Lifschitz equations, and explain how they are related to each other. In
section 3 we explain in general how specific choices for the gauge
transformations can become equivalent to Darboux transformations that when
iterated may be used to construct multi-soliton solutions. We then utilize
this scheme to derive explicit expressions for the multi-soliton solutions
to the continuous limit of the Heisenberg equation. These solutions may then
be used to calculate directly solutions to the Hirota equation as explained
in section 4. In section 5 we discuss how to obtain nonlocal multi-soliton
solutions to the extended Landau-Lifschitz equation. We provide two
alternative ways to achieve this, directly via the nonlocal solutions of the
Hirota equation or by implementing the nonlocality on some auxiliary
equations that emerge in the solution process of the continuous limit of the
Heisenberg equation. Our conclusions are stated in section 6.

\section{Nonlocal gauge equivalence}

Many integrable systems are related to each other by means of gauge
transformations, often in an unexpected way. Such type of correspondences
can be exploited to gain insight into either system from its gauge partner,
for instance by transforming solutions of one system to solutions of the
other. Often this process can only be carried out in one direction. One may
also consider auto-gauge transformation from a system to itself, which when
iterated can be used to generate new types of solutions, similar to Darboux
or auto-B\"{a}cklund transformations. In general, we consider here two
systems whose auxiliary functions $\Psi _{1}$ and $\Psi _{2}$ are related to
each other by means of a gauge field operator $G$ as $\Psi _{1}=G\Psi _{2}$.
Formally the system can be cast into two gauge equivalent zero curvature
conditions for the two sets of two operators, $U_{1}$, $V_{1}$ and $U_{2}$, $%
V_{2}$, together with their equivalent two linear first order differential
equations involving the auxiliary function $\Psi _{1}$ and $\Psi _{2}$ 
\begin{equation}
\partial _{t}U_{i}-\partial _{x}V_{i}+\left[ U_{i},V_{i}\right] =0\qquad
\Leftrightarrow \qquad \Psi _{i,t}=V_{i}\Psi _{i}\text{, }\Psi
_{i,x}=U_{i}\Psi _{i}~~~~i=1,2.  \label{ZC}
\end{equation}%
Given the transformation from $\Psi _{1}$ to $\Psi _{2}$, the operators $%
U_{1}$, $V_{1}$ and $U_{2}$, $V_{2}$ are related as%
\begin{equation}
U_{1}=GU_{2}G^{-1}+G_{x}G^{-1},\qquad \text{and\qquad }%
V_{1}=GV_{2}G^{-1}+G_{t}G^{-1}.  \label{U12}
\end{equation}%
The relations (\ref{ZC}) and (\ref{U12}) are entirely generic providing a
connection between two types of integrable systems, assuming the invertible
gauge transformation map $G$ exists. Specific systems are obtained by
concrete choices of the two sets of two operators $U_{1}$, $V_{1}$ and $%
U_{2} $, $V_{2}$.

\subsection{The nonlocal Hirota system}

Let us first specify the system $1$, by taking $U_{1}$ and $V_{1}$ to be of
the form \ 
\begin{equation}
U_{1}=A_{0}+\lambda A_{1},\qquad V_{1}=B_{0}+\lambda B_{1}+\lambda
^{2}B_{2}+\lambda ^{3}B_{3},\qquad  \label{U1}
\end{equation}%
where%
\begin{eqnarray}
A_{0} &=&\left( 
\begin{array}{cc}
0 & q(x,t) \\ 
r(x,t) & 0%
\end{array}%
\right) ,~~~A_{1}=\left( 
\begin{array}{cc}
-i & 0 \\ 
0 & i%
\end{array}%
\right) ~=-i\sigma _{3}, \\
B_{0} &=&i\alpha \left[ \sigma _{3}\left( A_{0}\right) _{x}-\sigma
_{3}A_{0}^{2}\right] +\beta \left[ 2A_{0}^{3}+\left( A_{0}\right)
_{x}A_{0}-A_{0}\left( A_{0}\right) _{x}-\left( A_{0}\right) _{xx}\right] , \\
B_{1} &=&2\alpha A_{0}+2i\beta \sigma _{3}\left[ \left( A_{0}\right)
_{x}-A_{0}^{2}\right] , \\
B_{2} &=&4\beta A_{0}-2i\alpha \sigma _{3}, \\
B_{3} &=&-4i\beta \sigma _{3},  \label{5}
\end{eqnarray}%
with $\sigma _{i}$, $i=1,2,3$ denoting the Pauli spin matrices, $\lambda $
the spectral parameter and $\alpha ,\beta $ are real constants. Using the
explicit expressions (\ref{U1})-(\ref{5}) the zero curvature condition (\ref%
{ZC}) becomes equivalent to the Hirota system \cite%
{hirota1973exact,CenFringHir} for the fields $q(x,t)$ and $r(x,t)$ as 
\begin{eqnarray}
q_{t}-i\alpha q_{xx}+2i\alpha q^{2}r+\beta \left[ q_{xxx}-6qrq_{x}\right]
&=&0,  \label{zero1} \\
r_{t}+i\alpha r_{xx}-2i\alpha qr^{2}+\beta \left( r_{xxx}-6qrr_{x}\right)
&=&0.  \label{zero2}
\end{eqnarray}%
These equations may be viewed with $q(x,t)$ and $r(x,t)$ as entirely
independent functions, but most commonly one imposes the relation $%
r(x,t)=q^{\ast }(x,t)$, such that the two equations\ become their mutual
conjugates and are therefore essentially reduced to one equation only - the
Hirota equation. Recently \cite{CenFringHir} alternative possibilities that
exploit $\mathcal{PT}$-symmetry have been proposed, such as taking $%
r(x,t)=\kappa q^{\ast }(-x,t)$, $r(x,t)=\kappa q^{\ast }(x,-t)$, $%
r(x,t)=\kappa q^{\ast }(-x,-t)$, $r(x,t)=\kappa q(-x,t)$, $r(x,t)=\kappa
q(x,-t)$ or $r(x,t)=\kappa q(-x,-t)$ with $\kappa \in \mathbb{R}$ and a
suitable adaptation of the parameters $\alpha $ and $\beta $. As for these
type of choices the equations contain fields that depend simultaneously on $%
x $ and $-x$, and/or $t$ and $-t$, they are referred to as \textit{nonlocal.}
These type of novel variants of integrable systems are the main focus of
this manuscript. It was shown in \cite{CenFringHir} that the different
versions display quite distinct and varied behaviour and therefore deserve
to be investigated in their own right. However, in what follows we will
exclusively focus on the complex parity extended version corresponding to
the choice $r(x,t)=\kappa q^{\ast }(-x,t)$ together with $\beta =i\delta $, $%
\delta \in \mathbb{R}$, and refer to it as the \textit{nonlocal} case
throughout the manuscript. The treatment of the other cases goes along the
same lines, but will not be discussed here.

\subsection{The nonlocal extended continuous Heisenberg equation}

Having committed to a fixed form of the system 1, we elaborate next on a
more precise form of the system $2$ following from that concrete choice.
Employing the expansion (\ref{U1}), we obtain from (\ref{U12}) the
expressions 
\begin{equation}
U_{2}=-i\lambda G^{-1}\sigma _{3}G,\qquad V_{2}=\lambda G^{-1}B_{1}G+\lambda
^{2}G^{-1}B_{2}G+\lambda ^{3}G^{-1}B_{3}G  \label{U2}
\end{equation}%
together with%
\begin{equation}
G_{x}=A_{0}G,\qquad \text{and\qquad }G_{t}=B_{0}G.  \label{10}
\end{equation}%
With given $A_{0}$ and $B_{0}$, it is the solution for these two equations
in (\ref{10}) that determines the precise form of the gauge map $G$ for a
particular set of models.

An interesting and universally applied equation emerges when we use the
gauge field $G$ to define a new field operator 
\begin{equation}
S:=G^{-1}\sigma _{3}G\text{.}  \label{SG}
\end{equation}%
The following properties follow directly from above 
\begin{equation}
S^{2}=1,\quad S_{x}=2G^{-1}\sigma _{3}A_{0}G,\quad
SS_{x}=-S_{x}S=2G^{-1}A_{0}G,\quad \left[ S,S_{xx}\right] =2\left(
SS_{xx}+S_{x}^{2}\right) .  \label{S}
\end{equation}%
Next we notice that instead of expressing the operators $U_{2}$ and $V_{2}$
in terms of the gauge field $G$, one can express them entirely in terms of
the operator $S$ as%
\begin{equation}
U_{2}=-i\lambda S,\qquad V_{2}=\alpha \left( \lambda SS_{x}-\lambda
^{2}2iS\right) +\beta \left[ \lambda \left( i\frac{3}{2}SS_{x}^{2}+iS_{xx}%
\right) +\lambda ^{2}2SS_{x}-\lambda ^{3}4iS\right] .  \label{UV2}
\end{equation}%
Using this variant we evaluate the zero curvature condition (\ref{ZC}) with (%
\ref{U2}) and the identities (\ref{S}), to obtain the equation of motion for
the $S$-operator%
\begin{eqnarray}
S_{t} &=&i\alpha \left( S_{x}^{2}+SS_{xx}\right) -\beta \left[ \frac{3}{2}%
\left( SS_{x}^{2}\right) _{x}+S_{xxx}\right]  \label{St} \\
&=&\frac{i}{2}\alpha \left[ S,S_{xx}\right] -\frac{\beta }{2}\left(
3S_{x}^{3}+S\left[ S,S_{xxx}\right] \right) .
\end{eqnarray}%
For $\beta =0$ this equation reduces to the well-known continuous limit of
the Heisenberg spin chain \cite%
{nakamura1974solitons,lakshmanan1,tjon1977solitons,takhtajan197,demontis2018}
and for $\beta \neq 0$ to the first member of the corresponding hierarchy 
\cite{wang2005darboux}. We refer to this equation as the extended continuous
Heisenberg (ECH) equation. The equation (\ref{St}) is rather universal as it
also emerges for other types of integrable higher order equations of
nonlinear Schr\"{o}dinger type, such as the modified Korteweg-de Vries
equation \cite{Kundu,MamKdV} or the Sasa--Satsuma equation \cite{Kundu,MaSS}%
. The distinction towards specific models of this general type is obtained
by specifying the gauge map $G$.

Given the above gauge correspondence one may now obtain solutions to the
nonlinear equations of a member of the nonlinear Schr\"{o}dinger hierarchy
from the equations of motion of the corresponding member the continuous
Heisenberg hierarchy, or vice versa. For instance, given a solution $%
q(x,t)\, $\ and $r(x,t)$ to the Hirota equations (\ref{zero1}), (\ref{zero2}%
) one may use equation (\ref{10}) to construct the gauge field operator $G$
and subsequently simply compute $S$, that solves (\ref{St}) by construction,
by means of the relation (\ref{SG}). In reverse, from a solution $S$ to (\ref%
{St}) we may construct $G$ by (\ref{SG}) and subsequently $q(x,t)\,$\ and $%
r(x,t)$ from (\ref{10}). We elaborate below on the details of this
correspondence.

\subsection{The nonlocal extended Landau-Lifschitz equation}

Equation (\ref{St}) possesses an interesting and well known vector variant
with many physical applications that arises when decomposing $S$ in the
standard fashion as $S=\mathbf{s\cdot \sigma }$, where $\mathbf{\sigma }$ is
a vector whose entries are Pauli matrices $\mathbf{\sigma =}(\sigma
_{1},\sigma _{2},\sigma _{3})$. Then the equation of motion (\ref{St})
becomes equivalent to an extended version of the Landau-Lifschitz equation%
\begin{equation}
\mathbf{s}_{t}=-\alpha \mathbf{s\times s}_{xx}-\frac{3}{2}\beta \left( 
\mathbf{s}_{x}\cdot \mathbf{s}_{x}\right) \mathbf{s}_{x}+\beta \mathbf{%
s\times }\left( \mathbf{s\times s}_{xxx}\right) .  \label{ELL}
\end{equation}%
The ELL equations (\ref{ELL}) is easily derived from (\ref{St}) when using
the standard identity $\sigma _{i}\sigma _{j}=\delta _{ij}\mathbb{I}%
+i\varepsilon _{ijk}\sigma _{k}$ with $\varepsilon _{ijk}$ denoting the
Levi-Civita tensor. Since $S^{2}=1$, we immediately obtain that $\mathbf{s}$
is a unit vector $\mathbf{s\cdot s}=1$. For $\beta \rightarrow 0$ this
equation reduces to the standard Landau-Lifschitz equation \cite%
{landau1935,bar2015landau}.

\section{Nonlocal multi-solitons for the ECHE from Darboux transformations}

\subsection{Darboux transformations as auto-gauge transformation}

Let us now explain how to solve the above nonlinear systems and construct
their nonlocal multi-soliton solutions by means of repeated gauge
transformations. The key idea is that the gauge is chosen in such a way that
the transformation becomes equivalent to a Darboux transformation. We start
with the extended continuous Heisenberg equation and introduce for
convenience $T:=-iS$, $\psi :=\Psi _{2}$, $U:=U_{2}~$and $V:=V_{2}$ so that
the spectral problem in (\ref{ZC}) for system $2$ with (\ref{UV2}) reads 
\begin{equation}
\psi _{x}=U\psi =\lambda T\psi ,\qquad \psi _{t}=V\psi
=\sum\limits_{k=0}^{2}\lambda ^{3-k}V^{(k)}\psi  \label{spp}
\end{equation}%
where%
\begin{equation}
V^{(0)}=4\beta T,\quad V^{(1)}=2\alpha T-2\beta TT_{x},\quad V^{(2)}=\frac{3%
}{2}\beta TT_{x}^{2}-\alpha TT_{x}-\beta T_{xx}.
\end{equation}%
Next we carry out another gauge transformation $\hat{G}$ on the system (\ref%
{spp}) relating the eigenstates $\psi $ to new eigenstates $\hat{\psi}$ as $%
\hat{\psi}=\hat{G}\psi $, so that similarly to the relations in (\ref{U12})
we obtain a new spectral problem with%
\begin{equation}
\hat{\psi}_{x}=\hat{U}\hat{\psi},\qquad \hat{\psi}_{t}=\hat{V}\hat{\psi},
\label{spec}
\end{equation}%
in which the new operators $\hat{U}$, $\hat{V}$ are related to the original $%
U$, $V$ as 
\begin{equation}
\hat{U}=\hat{G}U\hat{G}^{-1}+\hat{G}_{x}\hat{G}^{-1},\qquad \text{and\qquad }%
\hat{V}=\hat{G}V\hat{G}^{-1}+\hat{G}_{t}\hat{G}^{-1}.  \label{UVhat}
\end{equation}%
The key ingredient to achieve the equivalence between the gauge
transformation and the Darboux transformation \cite{darboux,matveevdarboux}
lies in the right choice of the gauge transformation $\hat{G}$. Following
essentially \cite{wang2005darboux}, we take now%
\begin{equation}
\hat{G}(\lambda ):=-\mathbb{I}+\lambda L\text{,~~~~\ with \ \ }L:=H\Lambda
^{-1}H^{-1}\text{, }H:=\left[ \psi (\lambda _{1}),\psi (\lambda _{2})\right] 
\text{, }\Lambda =\limfunc{diag}(\lambda _{1},\lambda _{2}).  \label{LH}
\end{equation}%
Thus $H$ is taken to be a $2\times 2$-matrix with column vectors $\psi
(\lambda _{1})$ and $\psi (\lambda _{2})$ denoting solution to (\ref{spp})
for some specific values of the spectral parameter $\lambda _{1}\neq \lambda
_{2}\neq 0$. We notice that $\det L=\lambda _{1}^{-1}\lambda _{2}^{-1}\neq 0$%
, so that the inverse of $L$ exists. Using (\ref{spp}) we then compute the
derivatives%
\begin{eqnarray}
H_{x} &=&TH\Lambda ,\quad \\
H_{t} &=&\sum\limits_{k=0}^{2}V^{(k)}H\Lambda ^{3-k},\quad \\
L_{x} &=&T-LTL^{-1}, \\
L_{t} &=&\sum\limits_{k=0}^{2}\left( V^{(k)}L^{k-2}-LV^{(k)}L^{k-3}\right) ,
\end{eqnarray}%
which allows us to evaluate the right hand sides of the equations in (\ref%
{UVhat}) to%
\begin{eqnarray}
\hat{U} &=&LUL^{-1},  \label{ut} \\
\hat{V}^{(0)} &=&LV^{(0)}L^{-1},  \label{v0} \\
\hat{V}^{(1)} &=&LV^{(1)}L^{-1}-V^{(0)}L^{-1}+\hat{V}^{(0)}L^{-1}, \\
\hat{V}^{(2)} &=&LV^{(2)}L^{-1}-V^{(1)}L^{-1}+\hat{V}^{(1)}L^{-1}.
\label{v3}
\end{eqnarray}%
The matrix $\hat{V}$ is of the same form as $V$, that is $\hat{V}%
=\sum\nolimits_{k=0}^{2}\lambda ^{3-k}\hat{V}^{(k)}$. Equation (\ref{ut}) is
equivalent to $\hat{S}L=LS$, which is reminiscent of the intertwining
relations employed in Darboux transformations. Since the gauge system is of
the same form as the original equation, this means that if $S$ is also a
solution to (\ref{St}), then $\hat{S}$ is a solution to the same equation.

We can now iterate this systems like a standard Darboux-Crum transformation 
\cite{darboux,crum}. Indexing all quantities, we have at each stage the
spectral problem%
\begin{equation}
\psi _{x}^{(n-1)}(\lambda )=U^{(n-1)}(\lambda )~\psi ^{(n-1)}(\lambda ),~~\
\ \psi _{t}^{(n-1)}(\lambda )=V^{(n-1)}(\lambda )~\psi ^{(n-1)}(\lambda
),~~~\ n\in \mathbb{N}\text{,}  \label{specn}
\end{equation}%
which when solved for $\psi ^{(n-1)}(\lambda )$ allows to define the new
quantities%
\begin{eqnarray}
H^{(n-1)}(\lambda _{2n-1},\lambda _{2n}) &:&=\left( \psi ^{(n-1)}(\lambda
_{2n-1}),\psi ^{(n-1)}(\lambda _{2n})\right) , \\
\Lambda _{n} &:&=\limfunc{diag}(\lambda _{2n-1},\lambda _{2n}), \\
L^{(n)}(\lambda _{2n-1},\lambda _{2n}) &:&=H^{(n-1)}(\lambda _{2n-1},\lambda
_{2n})\Lambda _{n}^{-1}\left[ H^{(n-1)}(\lambda _{2n-1},\lambda _{2n})\right]
^{-1},~~~  \label{LN}
\end{eqnarray}%
where $\lambda _{i}\neq \lambda _{j}\neq 0$, $i,j\in \mathbb{N}$. By means
of the intertwining operator $L^{(n)}$ we can now specify the gauge
transformations as%
\begin{equation}
\hat{G}^{(n)}(\lambda ):=-\mathbb{I}+\lambda L^{(n)},
\end{equation}%
so that we can construct the solution to the spectral problem (\ref{specn})
at the next level as%
\begin{eqnarray}
\psi ^{(n)}(\lambda ) &=&\hat{G}^{(n)}(\lambda )\psi ^{(n-1)}(\lambda )=%
\mathcal{G}^{(n)}(\lambda )\psi ^{(0)}(\lambda ),  \label{Pn} \\
U^{(n)}(\lambda ) &=&L^{(n)}U^{(n-1)}(\lambda )\left[ L^{(n)}\right] ^{-1}=%
\mathcal{L}^{(n)}U^{(0)}(\lambda )\left( \mathcal{L}^{(n)}\right) ^{-1},
\label{UU}
\end{eqnarray}%
with $\mathcal{L}^{(n)}:=L^{(n)}L^{(n-1)}\ldots L^{(1)}$ and $\mathcal{G}%
^{(n)}(\lambda ):=\hat{G}^{(n)}(\lambda )\hat{G}^{(n-1)}(\lambda )\ldots 
\hat{G}^{(1)}(\lambda )$. Noting that 
\begin{equation}
\det \mathcal{L}^{(n)}=\prod\nolimits_{i=1}^{2n}\lambda
_{i}^{-1}=(-1)^{n}\prod\nolimits_{i=1}^{n}\left\vert \lambda
_{2i-1}\right\vert ^{-2}=:\chi _{n}\text{,}
\end{equation}%
with $\lambda _{2i-1}=-\lambda _{2i}^{\ast }$, the inverse of $\mathcal{L}%
^{(n)}$ is guaranteed to always exist with the restrictions on the $\lambda
_{i}$ as introduced above. Extrapolating from (\ref{v0})-(\ref{v3}) there
are naturally also generic formulae for $V^{(n)}(\lambda )$, but since we
are mainly interested in $U^{(n)}(\lambda )$ we will not report them here.
It is clear from (\ref{specn})-(\ref{UU}) that once the initial spectral
problem involving $U^{(0)}(\lambda )$,$~V^{(0)}(\lambda )$ and $\psi
^{(0)}(\lambda )$ has been solved all higher levels follow simply by
iteration.

We apply this scheme now to construct multi-soliton solutions to equation (%
\ref{spp}) and in particular explain how nonlocality is naturally introduced
into these systems.

\subsection{Nonlocal multi-soliton solutions}

We start by parameterizing a matrix field solution $S$ to the extended
continuous Heisenberg equation (\ref{St}) as%
\begin{equation}
S=\left( 
\begin{array}{rr}
-\omega & u \\ 
v & \omega%
\end{array}%
\right) ,\qquad \omega ^{2}+uv=1,  \label{SS}
\end{equation}%
where the form of $S$ is dictated by (\ref{SG}) with the constraint on the
entries resulting from the first property in (\ref{S}). Substituting this
expression into equation (\ref{St}), we identify from the off-diagonal
components of this matrix equation the two constraining nonlinear
differential equations%
\begin{eqnarray}
u_{t} &=&i\alpha (u\omega _{x}-\omega u_{x})_{x}-\beta \left[
u_{xx}+3/2u(u_{x}v_{x}+\omega _{x}^{2})\right] _{x},  \label{uv1} \\
v_{t} &=&-i\alpha (v\omega _{x}-\omega v_{x})_{x}-\beta \left[
v_{xx}+3/2v(v_{x}u_{x}+\omega _{x}^{2})\right] _{x},  \label{uv2}
\end{eqnarray}%
that $u$, $v$ and $\omega $ have two satisfy. The diagonal entries are
trivially satisfied when (\ref{uv1}) and (\ref{uv2}) hold. Similarly as the
equations (\ref{zero1}) and (\ref{zero2}), one may treat (\ref{uv1}) and (%
\ref{uv2}) as independent equations for the functions $u(x,t)$ and $v(x,t)$,
with $\omega (x,t)$ obtained from the constraint in (\ref{SS}). However,
just as for the Hirota system one could also make the choice $u(x,t)=\kappa
v^{\ast }(x,t)$ so that equation (\ref{uv2}) simply becomes the complex
conjugate of equation (\ref{uv1}). Likewise we can make the nonlocal choice $%
u(x,t)=\kappa v^{\ast }(-x,t)$ with $\beta =i\delta $, $\delta \in \mathbb{R}
$, in which case equation (\ref{uv2}) becomes the complex conjugate parity
transformed of equation (\ref{uv1}). This means for the matrix $S$ that the
nonlocality can be imposed as $S(x,t)=\kappa S^{\dagger }(-x,t)$, which
holds with $\omega (x,t)=\kappa \omega ^{\ast }(-x,t)$.

The multi-soliton solutions to the ECH equation are then computed from the
Darboux-Crum transformations as explained above. From (\ref{UU}) we obtain
therefore 
\begin{equation}
S_{n}=\mathcal{L}^{(n)}S_{0}\left( \mathcal{L}^{(n)}\right) ^{-1},
\end{equation}%
with factors $L^{(i)}(\lambda _{2i-1},\lambda _{2i}),i=1,\ldots ,n$, being
evaluated with the appropriate two component solutions $(\psi (\lambda
_{2i-1}),\psi (\lambda _{2i}))$ to the spectral problems (\ref{specn}) for
the iterated $U$ and $V$ operators. As the entire procedure relies on the
solutions to the initial spectral problem, it is the choice of the so-called
seed functions $\psi ^{(0)}(\lambda _{i})=(\varphi _{i},\phi _{i})$ and
their implementation into the definition $H^{(n-1)}$ that will introduce the
nonlocality properties. This mechanism is similar to what we observed in 
\cite{CenFringHir}. From the iteration procedure we obtain the closed
solutions%
\begin{equation}
\mathcal{L}_{11}^{(n)}=\frac{\det \Omega _{n}}{\det \mathcal{W}_{n}},\qquad 
\mathcal{L}_{12}^{(n)}=\frac{\det \mathcal{U}_{n}}{\det \mathcal{W}_{n}}%
,\qquad \mathcal{L}_{21}^{(n)}=\frac{\det \mathcal{V}_{n}}{\det \mathcal{W}%
_{n}},\qquad \mathcal{L}_{22}^{(n)}=\frac{\det \Upsilon _{n}}{\det \mathcal{W%
}_{n}},  \label{Lgen}
\end{equation}%
with ($2n\times 2n$)-matrices $\mathcal{W}_{n}$, $\Omega _{n}$, $\Upsilon
_{n}$, $\mathcal{U}_{n}$ and $\mathcal{V}_{n}$ defined in terms of the seed
function components as%
\begin{equation}
\left( \mathcal{W}_{n}\right) _{ij}=\left\{ 
\begin{array}{l}
\lambda _{i}^{n+1-j}\varphi _{i} \\ 
\lambda _{i}^{2n+1-j}\phi _{i}%
\end{array}%
\right. \!\!,\left( \Omega _{n}\right) _{ij}=\left\{ 
\begin{array}{l}
\lambda _{i}^{j-1}\varphi _{i} \\ 
\lambda _{i}^{j-n}\phi _{i}%
\end{array}%
\right. \!\!,\left( \Upsilon _{n}\right) _{ij}=\left\{ 
\begin{array}{ll}
\lambda _{i}^{j}\varphi _{i}~ & j=1,\ldots ,n,~ \\ 
\lambda _{i}^{j-n-1}\phi _{i}~~~ & j=n+1,\ldots ,2n,%
\end{array}%
\right.
\end{equation}%
\begin{equation}
\left( \mathcal{U}_{n}\right) _{ij}=\left\{ 
\begin{array}{ll}
\lambda _{i}^{2n-j}\varphi _{i}~~~\ \  & j=n,\ldots ,2n, \\ 
\lambda _{i}^{n-j}\phi _{i} & j=1,\ldots ,n-1%
\end{array}%
\right. ,~\left( \mathcal{V}_{n}\right) _{ij}=\left\{ 
\begin{array}{ll}
\lambda _{i}^{j-1}\phi _{i}~~\ \  & j=1,\ldots ,n+1, \\ 
\lambda _{i}^{j-n-1}\varphi _{i} & j=n+2,\ldots ,2n,%
\end{array}%
\right.
\end{equation}%
with $i=1,\ldots ,2n$.

Keeping the matrix $S$ in the same functional form as in the
parameterization (\ref{SS}) at each step of the iteration procedure 
\begin{equation}
S_{n}=\left( 
\begin{array}{rr}
-\omega _{n} & u_{n} \\ 
v_{n} & \omega _{n}%
\end{array}%
\right) ,\qquad \omega _{n}^{2}+u_{n}v_{n}=1,  \label{Sn}
\end{equation}%
and abbreviating for convenience the entries of the matrix $\mathcal{L}%
^{(n)} $ by $A_{n}:=\mathcal{L}_{11}^{(n)}$, $B_{n}:=\mathcal{L}_{12}^{(n)}$%
, $C_{n}:=\mathcal{L}_{21}^{(n)}$, $D_{n}:=\mathcal{L}_{22}^{(n)}$ we
evaluate the entries to the $S$-matrix as 
\begin{eqnarray}
u_{n} &=&\left( A_{n}^{2}u_{0}-B_{n}^{2}v_{0}+2A_{n}B_{n}\omega _{0}\right)
/\chi _{n},  \label{uvw1} \\
v_{n} &=&\left( D_{n}^{2}v_{0}-C_{n}^{2}u_{0}-2C_{n}D_{n}\omega _{0}\right)
/\chi _{n},  \label{uvw2} \\
\omega _{n} &=&\left[
A_{n}C_{n}u_{0}-B_{n}D_{n}v_{0}+(A_{n}D_{n}+B_{n}C_{n})\omega _{0}\right]
/\chi _{n},  \label{uvw3}
\end{eqnarray}%
We also derive the identity 
\begin{equation}
\left( A_{n}\right) _{x}D_{n}-B_{n}\left( C_{n}\right) _{x}=A_{n}\left(
D_{n}\right) _{x}-\left( B_{n}\right) _{x}C_{n}=0,  \label{AB}
\end{equation}%
that will be crucial below. As mentioned, in order to obtain the nonlocal
solutions we need to impose the constraint $u_{n}(x,t)=\kappa v_{n}^{\ast
}(-x,t)$. Let us now explain how this is achieved by discussing the explicit
solutions in more detail.

\subsubsection{Nonlocal one-soliton solutions}

We start with a simple constant solution to the ECH equation (\ref{St}) of
the general form (\ref{Sn}) describing the free case%
\begin{equation}
S_{0}=\left( 
\begin{array}{rr}
-1 & 0 \\ 
0 & 1%
\end{array}%
\right) ,\qquad \omega _{0}=1\text{, }u_{0}=v_{0}=0.  \label{S0}
\end{equation}%
In order to define the matrix operator $H$ as in (\ref{LH}), we need to
construct the seed solution $\psi (\lambda )$ to the spectral problem (\ref%
{spp}) and evaluate it for two different and nonzero spectral parameters $%
\psi (\lambda _{1})=(\varphi _{1},\phi _{1})$ and $\psi (\lambda
_{2})=(\varphi _{2},\phi _{2})$. The first intertwining operator can be
computed directly and acquires the form%
\begin{equation}
\mathcal{L}^{(1)}=\frac{1}{\lambda _{1}\lambda _{2}\det H}\left( 
\begin{array}{cc}
\lambda _{2}\varphi _{1}\phi _{2}-\lambda _{1}\varphi _{2}\phi _{1} & 
(\lambda _{1}-\lambda _{2})\varphi _{1}\varphi _{2} \\ 
(\lambda _{2}-\lambda _{1})\phi _{1}\phi _{2} & \lambda _{1}\varphi _{1}\phi
_{2}-\lambda _{2}\varphi _{2}\phi _{1}%
\end{array}%
\right) .  \label{L1}
\end{equation}%
We confirm that this expression can be cast into the form of the generic
expression (\ref{Lgen}) with matrices%
\begin{equation}
\mathcal{W}_{1}=\left( 
\begin{array}{ll}
\lambda _{1}\varphi _{1} & \lambda _{1}\phi _{1} \\ 
\lambda _{2}\varphi _{2} & \lambda _{2}\phi _{2}%
\end{array}%
\right) ,~~~~~\Omega _{1}=\left( 
\begin{array}{ll}
\varphi _{1} & \lambda _{1}\phi _{1} \\ 
\varphi _{2} & \lambda _{2}\phi _{2}%
\end{array}%
\right) ,~~~~~\Upsilon _{1}=\left( 
\begin{array}{ll}
\lambda _{1}\varphi _{1} & \phi _{1} \\ 
\lambda _{2}\varphi _{2} & \phi _{2}%
\end{array}%
\right) ,~
\end{equation}%
\begin{equation}
\mathcal{U}_{1}=\left( 
\begin{array}{ll}
\lambda _{1}\varphi _{1} & \varphi _{1} \\ 
\lambda _{2}\varphi _{2} & \varphi _{2}%
\end{array}%
\right) ,~~~~~\mathcal{V}_{1}=\left( 
\begin{array}{ll}
\phi _{1} & \lambda _{1}\phi _{1} \\ 
\phi _{2} & \lambda _{2}\phi _{2}%
\end{array}%
\right) .
\end{equation}%
Given the intertwining operator $\mathcal{L}^{(1)}$, we can now calculate
the one-soliton solution $S_{1}$ directly from (\ref{uvw1})-(\ref{uvw3}),
obtaining%
\begin{eqnarray}
u_{1} &=&\frac{2\varphi _{1}\varphi _{2}(\lambda _{2}\varphi _{1}\phi
_{2}-\lambda _{1}\varphi _{2}\phi _{1})(\lambda _{1}-\lambda _{2})}{\lambda
_{1}\lambda _{2}(\varphi _{2}\phi _{1}-\varphi _{1}\phi _{2})^{2}},
\label{u1} \\
v_{1} &=&\frac{2\phi _{1}\phi _{2}(\lambda _{1}\varphi _{1}\phi _{2}-\lambda
_{2}\varphi _{2}\phi _{1})(\lambda _{1}-\lambda _{2})}{\lambda _{1}\lambda
_{2}(\varphi _{2}\phi _{1}-\varphi _{1}\phi _{2})^{2}},  \label{v1} \\
\omega _{1} &=&1-\frac{2\varphi _{1}\varphi _{2}\phi _{1}\phi _{2}(\lambda
_{1}-\lambda _{2})^{2}}{\lambda _{1}\lambda _{2}(\varphi _{2}\phi
_{1}-\varphi _{1}\phi _{2})^{2}}.~~  \label{w1}
\end{eqnarray}%
So far, these expressions hold for any solution to the spectral problem.
Imposing next the nonlocality condition $u_{1}(x,t)=\kappa v_{1}^{\ast
}(-x,t)$ leads for instance to the constraints%
\begin{equation}
\varphi _{2}(x,t)=-\kappa \phi _{1}^{\ast }(-x,t),\quad \phi
_{2}(x,t)=\varphi _{1}^{\ast }(-x,t),\quad \text{with }\lambda _{1}=-\lambda
_{2}^{\ast }=:\lambda \text{.}  \label{loco}
\end{equation}%
We can now solve the spectral problem (\ref{spp}) with $S_{0}$ for $\psi
(\lambda _{1}=\lambda )$ to%
\begin{equation}
\psi _{1}(\lambda )=\left( 
\begin{array}{l}
e^{\xi _{\lambda }(x,t)+\gamma _{1}} \\ 
e^{-\xi _{\lambda }(x,t)+\gamma _{2}}%
\end{array}%
\right) ,  \label{P1}
\end{equation}%
where we introduced the function%
\begin{equation}
\xi _{\lambda }(x,t):=i\lambda x+2\lambda ^{2}(i\alpha -2\delta \lambda )t.
\end{equation}%
and the additional constants $\gamma _{1},\gamma _{2}\in \mathbb{C}$ to
account for boundary conditions. The second solution is then simply obtained
from the constraint (\ref{loco}) to%
\begin{equation}
\psi _{2}(\lambda ^{\ast })=\left( 
\begin{array}{l}
-\kappa e^{-\xi _{\lambda }^{\ast }(-x,t)+\gamma _{2}^{\ast }} \\ 
e^{\xi _{\lambda }^{\ast }(-x,t)+\gamma _{1}^{\ast }}%
\end{array}%
\right) .  \label{P2}
\end{equation}%
Notice that $\psi _{2}(\lambda ^{\ast })$ is the solution to the parity
transformed and conjugated spectral problem (\ref{spp}). Given these
solutions we are in a position to compute the functions in (\ref{u1})-(\ref%
{w1})%
\begin{eqnarray}
u_{1}(x,t) &=&\frac{4\kappa \func{Re}\lambda \left( \kappa \lambda e^{-2\xi
_{\lambda }^{\ast }(-x,t)-\gamma _{1}^{\ast }+\gamma _{2}^{\ast }}-\lambda
^{\ast }e^{2\xi _{\lambda }(x,t)+\gamma _{1}-\gamma _{2}}\right) e^{2\func{Re%
}\gamma _{1}+2\func{Re}\gamma _{2}}}{\left\vert \lambda \right\vert
^{2}(e^{\xi _{\lambda }(x,t)+\xi _{\lambda }^{\ast }(-x,t)+2\func{Re}\gamma
_{1}}+\kappa e^{-\xi _{\lambda }(x,t)-\xi _{\lambda }^{\ast }(-x,t)+2\func{Re%
}\gamma _{2}})^{2}}, \\
v_{1}(x,t) &=&\frac{4\func{Re}\lambda \left( \kappa \lambda ^{\ast }e^{-2\xi
_{\lambda }(x,t)-\gamma _{1}+\gamma _{2}}-\lambda e^{2\xi _{\lambda }^{\ast
}(-x,t)+\gamma _{1}^{\ast }-\gamma _{2}^{\ast }}\right) e^{2\func{Re}\gamma
_{1}+2\func{Re}\gamma _{2}}}{\left\vert \lambda \right\vert ^{2}(e^{\xi
_{\lambda }(x,t)+\xi _{\lambda }^{\ast }(-x,t)+2\func{Re}\gamma _{1}}+\kappa
e^{-\xi _{\lambda }(x,t)-\xi _{\lambda }^{\ast }(-x,t)+2\func{Re}\gamma
_{2}})^{2}}, \\
\omega _{1}(x,t) &=&1-\frac{8\kappa (\func{Re}\lambda )^{2}e^{2\func{Re}%
\gamma _{1}+2\func{Re}\gamma _{2}}}{\left\vert \lambda \right\vert
^{2}(e^{\xi _{\lambda }(x,t)+\xi _{\lambda }^{\ast }(-x,t)+2\func{Re}\gamma
_{1}}+\kappa e^{-\xi _{\lambda }(x,t)-\xi _{\lambda }^{\ast }(-x,t)+2\func{Re%
}\gamma _{2}})^{2}}.~~
\end{eqnarray}%
We verify that these expressions do indeed satisfy the nonlinear
differential equations (\ref{uv1}) and (\ref{uv2}) for the component
functions of $S$, together with the locality constraint $u_{1}(x,t)=\kappa
v_{1}^{\ast }(-x,t)$ that converts the two equations (\ref{uv1}) and (\ref%
{uv2}) into each other via a parity transformation and a complex conjugation.

\subsubsection{Nonlocal two-soliton solutions}

The two-soliton solution is obtained in the next iterative step. With $%
L^{(1)}$ already computed in (\ref{L1}), we evaluate the gauge
transformation $\hat{G}^{(1)}$ and the matrix $H^{(1)}$ 
\begin{equation}
\hat{G}^{(1)}(\lambda )=-\mathbb{I}+\lambda L^{(1)},~~~H^{(1)}(\lambda
_{3},\lambda _{4}):=\left( \hat{G}^{(1)}(\lambda _{3})\psi ^{(0)}(\lambda
_{3}),\hat{G}^{(1)}(\lambda _{4})\psi ^{(0)}(\lambda _{4})\right) ,
\end{equation}%
from which we compute $L^{(2)}(\lambda _{3},\lambda _{4})$ as defined in (%
\ref{LN}). Subsequently we compute the complete intertwining operator $%
\mathcal{L}^{(2)}=L^{(2)}(\lambda _{3},\lambda _{4})L^{(1)}(\lambda
_{1},\lambda _{2})$ with entries given by the general formula (\ref{Lgen})
with explicit matrices%
\begin{equation}
\mathcal{W}_{2}=\left( 
\begin{array}{cccc}
\lambda _{1}^{2}\varphi _{1} & \lambda _{1}\varphi _{1} & \lambda
_{1}^{2}\phi _{1} & \lambda _{1}\phi _{1} \\ 
\lambda _{2}^{2}\varphi _{2} & \lambda _{2}\varphi _{2} & \lambda
_{2}^{2}\phi _{2} & \lambda _{2}\phi _{2} \\ 
\lambda _{3}^{2}\varphi _{3} & \lambda _{3}\varphi _{3} & \lambda
_{3}^{2}\phi _{3} & \lambda _{3}\phi _{3} \\ 
\lambda _{4}^{2}\varphi _{4} & \lambda _{4}\varphi _{4} & \lambda
_{4}^{2}\phi _{4} & \lambda _{4}\phi _{4}%
\end{array}%
\right) ,
\end{equation}%
\begin{equation}
~\Omega _{2}=\left( 
\begin{array}{cccc}
\varphi _{1} & \lambda _{1}\varphi _{1} & \lambda _{1}\phi _{1} & \lambda
_{1}^{2}\phi _{1} \\ 
\varphi _{2} & \lambda _{2}\varphi _{2} & \lambda _{2}\phi _{2} & \lambda
_{2}^{2}\phi _{2} \\ 
\varphi _{3} & \lambda _{3}\varphi _{3} & \lambda _{3}\phi _{3} & \lambda
_{3}^{2}\phi _{3} \\ 
\varphi _{4} & \lambda _{4}\varphi _{4} & \lambda _{4}\phi _{4} & \lambda
_{4}^{2}\phi _{4}%
\end{array}%
\right) ,~~~~~\Upsilon _{2}=\left( 
\begin{array}{cccc}
\lambda _{1}\varphi _{1} & \lambda _{1}^{2}\varphi _{1} & \phi _{1} & 
\lambda _{1}\phi _{1} \\ 
\lambda _{2}\varphi _{2} & \lambda _{2}^{2}\varphi _{2} & \phi _{2} & 
\lambda _{2}\phi _{2} \\ 
\lambda _{3}\varphi _{3} & \lambda _{3}^{2}\varphi _{3} & \phi _{3} & 
\lambda _{3}\phi _{3} \\ 
\lambda _{4}\varphi _{4} & \lambda _{4}^{2}\varphi _{4} & \phi _{4} & 
\lambda _{4}\phi _{4}%
\end{array}%
\right) ,~
\end{equation}%
\begin{equation}
\mathcal{U}_{2}=\left( 
\begin{array}{cccc}
\lambda _{1}\phi _{1} & \lambda _{1}^{2}\varphi _{1} & \lambda _{1}\varphi
_{1} & \varphi _{1} \\ 
\lambda _{2}\phi _{2} & \lambda _{2}^{2}\varphi _{2} & \lambda _{2}\varphi
_{2} & \varphi _{2} \\ 
\lambda _{3}\phi _{3} & \lambda _{3}^{2}\varphi _{3} & \lambda _{3}\varphi
_{3} & \varphi _{3} \\ 
\lambda _{4}\phi _{4} & \lambda _{4}^{2}\varphi _{4} & \lambda _{4}\varphi
_{4} & \varphi _{4}%
\end{array}%
\right) ~,~~~~\mathcal{V}_{2}=\left( 
\begin{array}{cccc}
\phi _{1} & \lambda _{1}\phi _{1} & \lambda _{1}^{2}\phi _{1} & \lambda
_{1}\varphi _{1} \\ 
\phi _{2} & \lambda _{2}\phi _{2} & \lambda _{2}^{2}\phi _{2} & \lambda
_{2}\varphi _{2} \\ 
\phi _{3} & \lambda _{3}\phi _{3} & \lambda _{3}^{2}\phi _{3} & \lambda
_{3}\varphi _{3} \\ 
\phi _{4} & \lambda _{4}\phi _{4} & \lambda _{4}^{2}\phi _{4} & \lambda
_{4}\varphi _{4}%
\end{array}%
\right) .
\end{equation}%
For the nonlocal case we define $\psi _{1}$ and $\psi _{2}$ as in (\ref{P1})
and (\ref{P2}). In addition, we use%
\begin{equation}
\psi _{3}(\lambda _{3}=\rho )=\left( 
\begin{array}{l}
\varphi _{3} \\ 
\phi _{3}%
\end{array}%
\right) =\left( 
\begin{array}{l}
e^{\xi _{\rho }(x,t)+\gamma _{3}} \\ 
e^{-\xi _{\rho }(x,t)+\gamma _{4}}%
\end{array}%
\right) ,~~~~\psi _{4}(\lambda _{4}=\rho ^{\ast })=\left( 
\begin{array}{l}
\varphi _{4} \\ 
\phi _{4}%
\end{array}%
\right) =\left( 
\begin{array}{l}
-\kappa e^{-\xi _{\rho }^{\ast }(-x,t)+\gamma _{4}^{\ast }} \\ 
e^{\xi _{\rho }^{\ast }(-x,t)+\gamma _{3}^{\ast }}%
\end{array}%
\right) ,
\end{equation}%
so that with (\ref{uvw1})-(\ref{uvw3}) we determine the nonlocal two-soliton
solutions as%
\begin{eqnarray}
\omega _{2} &=&\sqrt{1-u_{2}(x,t)v_{2}(x,t)},  \label{2s} \\
u_{2} &=&\frac{2\left( L_{1234}-L_{2134}+L_{3124}-L_{4123}\right) \left(
R_{1234}+R_{1342}+R_{1423}+R_{2314}+R_{2431}+R_{3412}\right) }{\lambda
_{1}\lambda _{2}\lambda _{3}\lambda _{4}\left( \Gamma _{1234}+\Gamma
_{1342}+\Gamma _{1423}+\Gamma _{2314}+\Gamma _{3412}+\Gamma _{4213}\right)
{}^{2}},  \notag \\
v_{2} &=&\frac{2\left( K_{2134}-K_{1234}+K_{4123}-K_{3124}\right) \left(
T_{1234}+T_{1342}+T_{1423}+T_{2314}+T_{2431}+T_{3412}\right) }{\lambda
_{1}\lambda _{2}\lambda _{3}\lambda _{4}\left( \Gamma _{1234}+\Gamma
_{1342}+\Gamma _{1423}+\Gamma _{2314}+\Gamma _{3412}+\Gamma _{4213}\right)
{}^{2}},  \notag
\end{eqnarray}%
where we defined the shorthand symbols 
\begin{eqnarray}
\Gamma _{ijkl} &:&=(\lambda _{i}-\lambda _{j})(\lambda _{k}-\lambda
_{l})\varphi _{i}\varphi _{j}\phi _{k}\phi _{l},~~R_{ijkl}=\lambda
_{k}\lambda _{l}\Gamma _{ijkl},~~T_{ijkl}=\lambda _{i}\lambda _{j}\Gamma
_{ijkl}, \\
L_{ijkl} &:&=\lambda _{i}(\lambda _{j}-\lambda _{k})(\lambda _{j}-\lambda
_{l})(\lambda _{k}-\lambda _{l})\phi _{i}\varphi _{j}\varphi _{k}\varphi
_{l},~~ \\
K_{ijkl} &:&=\lambda _{i}(\lambda _{j}-\lambda _{k})(\lambda _{j}-\lambda
_{l})(\lambda _{k}-\lambda _{l})\varphi _{i}\phi _{j}\phi _{k}\phi _{l},
\end{eqnarray}%
Once more we verify that these expressions satisfy the nonlinear
differential equations (\ref{uv1}) and (\ref{uv2}) for the component
functions of $S$ and in addition are nonlocal, i.e. satisfying $%
u_{2}(x,t)=\kappa v_{2}^{\ast }(-x,t)$, which is required to convert the two
equations into each other via a parity transformation and a complex
conjugation.

\subsection{Nonlocal $n$-soliton solutions}

We proceed further in the same way for the nonlocal multi-soliton solutions
for $n>2$. In general, for a nonlocal $n$-soliton solution we choose the
spectral parameters as 
\begin{equation}
\lambda _{2i}=-\lambda _{2i-1}^{\ast }\neq 0,~~~~\ \lambda _{2i}\neq \lambda
_{2j}\ \ \ i,j=1,2,\ldots ,n,
\end{equation}%
and the seed functions computed at these values as%
\begin{eqnarray}
\psi _{2i-1}(\lambda _{2i-1}) &=&\left( 
\begin{array}{l}
\varphi _{2i-1} \\ 
\phi _{2i-1}%
\end{array}%
\right) =\left( 
\begin{array}{l}
e^{\xi _{\lambda _{2i-1}}(x,t)+\gamma _{2i-1}} \\ 
e^{-\xi _{\lambda _{2i-1}}(x,t)+\gamma _{2i}}%
\end{array}%
\right) ,~~~~~~  \label{seed1} \\
\psi _{2i}(\lambda _{2i}) &=&\left( 
\begin{array}{l}
\varphi _{2i} \\ 
\phi _{2i}%
\end{array}%
\right) =\left( 
\begin{array}{l}
-\kappa e^{-\xi _{\lambda _{2i-1}}^{\ast }(-x,t)+\gamma _{2i}^{\ast }} \\ 
e^{\xi _{\lambda _{2i-1}}^{\ast }(-x,t)+\gamma _{2i-1}^{\ast }}%
\end{array}%
\right) .  \label{seed2}
\end{eqnarray}%
We may then apply directly the formulae (\ref{uvw1})-(\ref{uvw3}) and
evaluate $u_{n}$, $v_{n}$, and $\omega _{n}$. We find the nonlocality
property $u_{n}(x,t)=\kappa v_{n}^{\ast }(-x,t)$ for all solutions.

\section{Nonlocal solutions to Hirota's equation from the ECH equation}

Let us now demonstrate how to obtain nonlocal solutions for the Hirota
equation from those of the extended continuous Heisenberg equation. With $S$
being parameterized as in (\ref{SS}) we solve for this purpose equation (\ref%
{SG}) for $G$ 
\begin{equation}
G=\left( 
\begin{array}{rr}
a & a\frac{\omega +1}{v} \\ 
c & c\frac{\omega -1}{v}%
\end{array}%
\right) ,
\end{equation}%
where the functions $a(x,t)$ and $c(x,t)$ remain unknown at this point. They
can be determined when substituting $G$ into the equations (\ref{10}).
Solving the first equation for $q(x,t)\,$and $r(x,t)$ we find 
\begin{eqnarray}
q(x,t) &=&\frac{\mu (t)}{2}\left( \frac{v_{x}}{v}+\frac{\omega v_{x}-\omega
_{x}v}{v}\right) \exp \left[ \int^{x}\frac{\omega (s,t)v_{s}(s,t)-\omega
_{s}(s,t)v(s,t)}{v(s,t)}ds\right] ,  \label{qint} \\
r(x,t) &=&\frac{1}{2\mu (t)}\left( \frac{v_{x}}{v}-\frac{\omega v_{x}-\omega
_{x}v}{v}\right) \exp \left[ -\int^{x}\frac{\omega (s,t)v_{s}(s,t)-\omega
_{s}(s,t)v(s,t)}{v(s,t)}ds\right] ,~~~  \label{rint}
\end{eqnarray}%
where $\mu (t)$ is an arbitrary function of $t$ at this stage. Notice that
the integral representations (\ref{qint}) and (\ref{rint}) are valid for 
\textit{any }solution to the ECH equation (\ref{St}). Next we demonstrate
how to solve these integrals. Using the expression in (\ref{uvw1})-(\ref%
{uvw3}), with suppressed subscripts $n$ and $S_{0}$ chosen as in (\ref{SS}),
we can re-express the terms in (\ref{qint}) and (\ref{rint}) via the
components of the intertwining operator $\mathcal{L}$ as 
\begin{eqnarray}
\frac{\omega v_{x}-\omega _{x}v}{v} &=&-2\frac{A_{x}D-BC_{x}}{AD-BC}%
+\partial _{x}\ln \left[ \frac{C}{D}\left( AD-BC\right) \right] =\partial
_{x}\ln \left( \frac{C}{D}\right) , \\
\frac{v_{x}}{v} &=&\frac{D}{C}\frac{AC_{x}-A_{x}C}{AD-BC}-\frac{C}{D}\frac{%
BD_{x}-B_{x}D}{AD-BC}=\frac{C_{x}}{C}+\frac{D_{x}}{D},
\end{eqnarray}%
where we used the property (\ref{AB}). With these relations the integral
representations (\ref{qint}), (\ref{rint}) simplify to 
\begin{eqnarray}
q_{n}(x,t) &=&c\mu _{n}\frac{\left( C_{n}\right) _{x}}{D_{n}}=c\mu
_{n}\left( \frac{\left( \det \mathcal{V}_{n}\right) _{x}}{\det \Upsilon _{n}}%
-\frac{\left( \det \mathcal{W}_{n}\right) _{x}\det \mathcal{V}_{n}}{\det 
\mathcal{W}_{n}\det \Upsilon _{n}}\right) ,  \label{qn} \\
r_{n}(x,t) &=&\frac{1}{c\mu _{n}}\frac{\left( D_{n}\right) _{x}}{C_{n}}=%
\frac{1}{c\mu _{n}}\left( \frac{\left( \det \Upsilon _{n}\right) _{x}}{\det 
\mathcal{V}_{n}}-\frac{\left( \det \mathcal{W}_{n}\right) _{x}\det \Upsilon
_{n}}{\det \mathcal{W}_{n}\det \mathcal{V}_{n}}\right) ,~~~  \label{rn}
\end{eqnarray}%
where $c$ is an integration constant. Thus, we have now obtained a simple
relation between the spectral problem of the extended continuous Heisenberg
equation and the solutions to the Hirota equation. It appears that this is a
novel relation even for the local scenario. The nonlocality property of the
solutions to the ECH equation is then naturally inherited by the solutions
to the Hirota equation. Using the nonlocal choices for the seed functions as
specified in (\ref{seed1}) and (\ref{seed2}) we may compute directly the
right hand sides in (\ref{qn}) and (\ref{rn}). Crucially these solutions
satisfy the nonlocality property%
\begin{equation}
r_{n}(x,t)=\frac{\kappa }{c^{2}\mu _{n}^{2}}q_{n}^{\ast }(-x,t).
\end{equation}%
We discuss this in more detail for the one-soliton solution for which the
more explicit expressions are less lengthy.

\subsection{Nonlocal one-soliton solutions}

Reading off the entries $C_{1}$ and $D_{1}$ from the $\mathcal{L}^{(1)}$%
-operator in (\ref{L1}), the one-soliton solution in (\ref{qn}) and (\ref{rn}%
) acquires the form%
\begin{eqnarray}
q_{1}(x,t) &=&c\mu _{1}(\text{$\lambda $}_{1}-\text{$\lambda _{2}$})\frac{%
\text{$\phi $}_{1}^{2}\left[ \varphi _{2}(\text{$\phi $}\text{$_{2}$})_{x}-%
\text{$\phi $}\text{$_{2}$}\left( \varphi _{2}\right) _{x}\right] +\text{$%
\phi $}_{2}^{2}\left[ \text{$\phi $}\text{$_{1}$}\left( \varphi _{1}\right)
_{x}-\varphi _{1}(\text{$\phi $}\text{$_{1}$})_{x}\right] }{(\varphi _{2}%
\text{$\phi $}\text{$_{1}$}-\varphi _{1}\text{$\phi $}\text{$_{2}$})(\lambda
_{1}\varphi _{1}\text{$\phi $}\text{$_{2}$}-\text{$\lambda $}_{2}\varphi _{2}%
\text{$\phi $}\text{$_{1}$})}, \\
r_{1}(x,t) &=&\frac{1}{c\mu _{1}}\frac{\phi _{1}\phi _{2}\left[ \varphi
_{2}\left( \varphi _{1}\right) _{x}-\varphi _{1}\left( \varphi _{2}\right)
_{x}\right] -\varphi _{1}\varphi _{2}\left[ \phi _{2}\left( \phi _{1}\right)
_{x}+\phi _{1}\left( \phi _{2}\right) _{x}\right] }{\phi _{1}\phi _{2}\left(
\varphi _{2}\phi _{1}-\varphi _{1}\phi _{2}\right) }.
\end{eqnarray}%
Specifying the solutions to the spectral problem as in (\ref{P1}) and (\ref%
{P2}) with $\lambda $$_{1}=\lambda $, $\lambda $$_{2}=-\lambda ^{\ast }$ and 
$c=-1$ we obtain the nonlocal one-soliton solution 
\begin{eqnarray}
q_{1}(x,t) &=&4i\mu _{1}\func{Re}\lambda \frac{e^{-\xi (x,t)+\text{$\xi $}%
^{\ast }(-x,t)+\gamma _{2}+\gamma _{1}^{\ast }}}{e^{\xi (x,t)+\text{$\xi $}%
^{\ast }(-x,t)+\gamma _{1}+\gamma _{1}^{\ast }}+\kappa e^{-\xi (x,t)-\text{$%
\xi $}^{\ast }(-x,t)+\gamma _{2}+\gamma _{2}^{\ast }}},  \label{s1} \\
r_{1}(x,t) &=&-\frac{4i\kappa \func{Re}\lambda }{\mu _{1}}\frac{e^{\gamma
_{1}+\gamma _{2}^{\ast }}}{e^{2\text{$\xi $}^{\ast }(-x,t)+\gamma
_{1}+\gamma _{1}^{\ast }}+\kappa e^{-2\xi (x,t)+\gamma _{2}+\gamma
_{2}^{\ast }}}.  \label{s2}
\end{eqnarray}%
We verify that (\ref{s1}) and (\ref{s2}) are solution to the nonlocal Hirota
equation (\ref{zero1}) and (\ref{zero2}) for any constant value of $\mu _{1}$%
. The nonlocality property inherited from the extended continuous Heisenberg
equation is%
\begin{equation}
r_{1}(x,t)=\frac{\kappa }{\mu _{1}^{2}}q_{1}^{\ast }(-x,t).  \label{nonloc}
\end{equation}%
Thus for $\mu _{1}^{2}=1$ the nonlocality property between $q_{1}$ and $r_{1}
$ becomes the same as the one between the functions $v_{1}$ and $u_{1}$.
Clearly, we may proceed in the same manner and use (\ref{qn}) and (\ref{rn})
to calculate directly the nonlocal $n$-soliton solutions to the to the
Hirota equation from the spectral problem of the nonlocal ECH equation.

\section{Nonlocal solutions to the extended Landau-Lifschitz equation}

\subsection{Nonlocal solutions to the ELLE from the ECHE or Hirota equation}

Given the nonlocal solutions to the ECH equation (\ref{St}), it is now also
straightforward to construct nonlocal solutions\ to the ELL equation (\ref%
{ELL}) from them simply by using the representation $S_{n}=\mathbf{s}_{n}%
\mathbf{\cdot \sigma }$ with $S_{n}$ taken to be in the parameterization (%
\ref{Sn}). Suppressing the index $n$, a direct expansion then yields%
\begin{equation}
s_{1}=\frac{1}{2}(u+v),\qquad s_{2}=\frac{i}{2}(u-v),\qquad s_{3}=-\omega
=\pm \sqrt{1-uv}.  \label{s123}
\end{equation}%
For the local choice $u(x,t)=v^{\ast }(x,t)$ these function are evidently
real%
\begin{equation}
s_{1}(x,t)=\func{Re}u,\qquad s_{2}=-\func{Im}u,\qquad s_{3}=\pm \sqrt{%
1-\left\vert u\right\vert ^{2}}.
\end{equation}%
Thus, since $\mathbf{s}$ is a real unit vector function and $\mathbf{s\cdot s%
}=1$, its endpoint traces out a curve on the unit sphere, as demonstrated
with an example of a one-soliton solution in figure 1 below. However, for
the nonlocal choice $u(x,t)=\kappa v^{\ast }(-x,t)$, the vector function $%
\mathbf{s}$ is no longer real so that we may decompose it into $\mathbf{s}=%
\mathbf{m}+i\mathbf{l}$, where now $\mathbf{m}$ and $\mathbf{l}$ are real
valued vector functions. From the relation $\mathbf{s\cdot s}=1$ it follows
directly that $\mathbf{m}^{2}-$ $\mathbf{l}^{2}=1~$and that these vector
functions are orthogonal to each other $\mathbf{m}\cdot \mathbf{l}=0$. The
nonlocal extended Landau Lifschitz equation (\ref{ELL}) then becomes a set
of coupled equations for the real valued vector functions $\mathbf{m}$ and $%
\mathbf{l}$ 
\begin{eqnarray}
\mathbf{m}_{t} &=&\alpha \left( \mathbf{l\times l}_{xx}-\mathbf{m\times m}%
_{xx}\right) +\frac{3}{2}\delta \left[ \left( \mathbf{m}_{x}\cdot \mathbf{m}%
_{x}\right) \mathbf{m}_{x}+2\left( \mathbf{l}_{x}\cdot \mathbf{m}_{x}\right) 
\mathbf{m}_{x}-\left( \mathbf{l}_{x}\cdot \mathbf{l}_{x}\right) \mathbf{l}%
_{x}\right] ~~~~~~~  \label{NELL1} \\
&&+\delta \left[ \mathbf{l\times }\left( \mathbf{l\times l}_{xxx}\right) -%
\mathbf{m\times }\left( \mathbf{l\times m}_{xxx}\right) -\mathbf{m\times }%
\left( \mathbf{m\times l}_{xxx}\right) -\mathbf{l\times }\left( \mathbf{%
m\times m}_{xxx}\right) \right] ,  \notag \\
\mathbf{l}_{t} &=&-\alpha \left( \mathbf{l\times m}_{xx}+\mathbf{m\times l}%
_{xx}\right) +\frac{3}{2}\delta \left[ \left( \mathbf{l}_{x}\cdot \mathbf{l}%
_{x}\right) \mathbf{m}_{x}+2\left( \mathbf{l}_{x}\cdot \mathbf{m}_{x}\right) 
\mathbf{l}_{x}-\left( \mathbf{m}_{x}\cdot \mathbf{m}_{x}\right) \mathbf{m}%
_{x}\right]  \label{NELL2} \\
&&+\delta \left[ \mathbf{m\times }\left( \mathbf{m\times m}_{xxx}\right) -%
\mathbf{l\times }\left( \mathbf{m\times l}_{xxx}\right) -\mathbf{l\times }%
\left( \mathbf{l\times m}_{xxx}\right) -\mathbf{m\times }\left( \mathbf{%
l\times l}_{xxx}\right) \right] ,  \notag
\end{eqnarray}%
Given $\mathbf{s}$, the real component entries of the new vectors are
trivially obtained from (\ref{s123}) to $m_{i}=\left( s_{i}+s_{i}^{\ast
}\right) /2$ and $l_{i}=i\left( s_{i}^{\ast }-s_{i}\right) /2$ that is%
\begin{eqnarray}
m_{1}(x,t) &=&\frac{1}{4}\left[ u(x,t)+v(x,t)+\kappa v(-x,t)+\kappa
^{-1}u(-x,t)\right] ,  \label{m1} \\
m_{2}(x,t) &=&\frac{i}{4}\left[ u(x,t)-v(x,t)-\kappa v(-x,t)+\kappa
^{-1}u(-x,t)\right] , \\
m_{3}(x,t) &=&-\frac{1}{2}\left[ \omega (x,t)+\omega (-x,t)\right] \\
l_{1}(x,t) &=&\frac{i}{4}\left[ -u(x,t)-v(x,t)+\kappa v(-x,t)+\kappa
^{-1}u(-x,t)\right] , \\
l_{2}(x,t) &=&\frac{1}{4}\left[ u(x,t)-v(x,t)+\kappa v(-x,t)-\kappa
^{-1}u(-x,t)\right] , \\
l_{3}(x,t) &=&\frac{i}{2}\left[ \omega (x,t)-\omega (-x,t)\right] .
\label{l3}
\end{eqnarray}%
Clearly despite the fact that $\mathbf{s\cdot s}=1$, the real and imaginary
components no longer trace out a curve on the unit sphere.

When solving the ECH equation directly we have implemented to nonlocality
through the compatibility relations between the auxiliary equations (\ref%
{uv1}) and (\ref{uv2}), which was then inherited by $\mathbf{s}$. We may
also attempt to implement the nonlocality from the Hirota system directly
into $S$ and therefore $\mathbf{s}$. For this purpose we make use of the
fact that so far the gauge operator $G$, that relates the spectral problem
of the Hirota system to the spectral problem of the ECH equation has been
left completely generic and the entries of the matrix $A_{0}$ are
constrained by the equations of motion (\ref{zero1}) and (\ref{zero2}).

As commented above, when specifying the $(A_{0})_{21}$-entry to $%
r(x,t)=\kappa q^{\ast }(x,t)$ with $\kappa =\pm 1$, the two equations (\ref%
{zero1}) and (\ref{zero2}) reduce to standard local Hirota equation \cite%
{hirota1973exact,CenFringHir}. The first equation in (\ref{10}) then implies
that $G_{11}=G_{22}^{\ast }$ and $\ G_{21}=\kappa G_{12}^{\ast }$, reducing
the four equations resulting from each matrix entry to the two equations%
\begin{equation}
a_{x}=\kappa b^{\ast }u,\qquad b_{x}=b^{\ast }u,  \label{11}
\end{equation}%
where we used the more compact notation $G_{11}=:a$, $G_{12}=:b$. Having
specified the gauge transformation $G$, we may compute the matrix $S$
directly from its defining relation (\ref{SG}) so that the components of the
vector $\mathbf{s}$ become in this case%
\begin{equation}
s_{1}=\frac{a^{\ast }b-\kappa ab^{\ast }}{\left\vert a\right\vert
^{2}-\kappa \left\vert b\right\vert ^{2}},\quad s_{2}=i\frac{a^{\ast
}b+\kappa ab^{\ast }}{\left\vert a\right\vert ^{2}-\kappa \left\vert
b\right\vert ^{2}},\quad s_{3}=\frac{\left\vert a\right\vert ^{2}+\kappa
\left\vert b\right\vert ^{2}}{\left\vert a\right\vert ^{2}-\kappa \left\vert
b\right\vert ^{2}}.
\end{equation}%
Hence for the choice $\kappa =-1$ the vector $\mathbf{s}$ is real valued.

For the nonlocal choice $r(x,t)=\kappa q^{\ast }(-x,t)$ the first equation
in (\ref{10}) implies that $G_{11}=\tilde{G}_{22}^{\ast }$ and $%
G_{21}=-\kappa \tilde{G}_{12}^{\ast }$. We adopt here the notation from \cite%
{CenFringHir} and suppress the explicit dependence on $(x,t)$, indicating
the functional dependence on $(-x,t)$ by a tilde, i.e. $\tilde{q}:=q(-x,t)$, 
$\tilde{G}_{12}^{\ast }:=G_{12}^{\ast }(-x,t)$, etc. The first equation in (%
\ref{10}) then reduces to the two equations%
\begin{equation}
a_{x}=-\kappa \tilde{b}^{\ast }u,\qquad b_{x}=\tilde{a}^{\ast }u,
\label{nl1}
\end{equation}%
so that in this case the components of the vector $\mathbf{s}$ become%
\begin{equation}
s_{1}=\frac{\tilde{a}^{\ast }b-\kappa a\tilde{b}^{\ast }}{a\tilde{a}^{\ast
}-\kappa b\tilde{b}^{\ast }},\quad s_{2}=i\frac{\tilde{a}b+\kappa a\tilde{b}%
^{\ast }}{a\tilde{a}^{\ast }-\kappa b\tilde{b}^{\ast }},\quad s_{3}=i\frac{a%
\tilde{a}^{\ast }+\kappa b\tilde{b}^{\ast }}{a\tilde{a}^{\ast }-\kappa b%
\tilde{b}^{\ast }},  \label{sss}
\end{equation}%
which are solutions to the nonlocal extended Landau-Lifschitz equations (\ref%
{NELL1}), (\ref{NELL2}).

\subsection{Nonlocal one-soliton solutions to the ELL equation}

We will now discuss some concrete soliton solutions obtained as outlined in
the previous section. We start by solving the constraint (\ref{11}) for a
local solution first that determines the gauge transformation $G$. Making
the additional assumption $b=cq(x,t)$ for some constant $c$, equation (\ref%
{11}) becomes%
\begin{equation}
a_{x}=\kappa c\left\vert q\right\vert ^{2},\qquad a=c\left( \ln q^{\ast
}\right) _{x}.  \label{PP}
\end{equation}%
The compatibility between the two equations in (\ref{PP}) implies that%
\begin{equation}
\left\vert q\right\vert ^{2}=\kappa \left( \ln q^{\ast }\right) _{xx}\quad
\Leftrightarrow \quad D_{x}^{2}f\cdot f=-2\kappa \left\vert g\right\vert ^{2}%
\text{,~~}\left( \ln g\right) _{xx}=0\text{, }  \label{co}
\end{equation}%
with $q=g/f$ , $g\in \mathbb{C}$, $f\in \mathbb{R}$. We notice that the
first relation in (\ref{co}), following directly from (\ref{PP}),
corresponds to one of the bilinear equations into which the Hirota equation
can be converted with an additional constraint. Here $D_{x}$ denotes a
Hirota derivative, see \cite{CenFringHir} for more details. Evidently the
additional constraint is not satisfied by all solutions to the Hirota
equation. The second equation in (\ref{10}) is then satisfied trivially for
solutions of (\ref{co}). We have therefore obtained a solution for the gauge
field operator in the form%
\begin{equation}
G=c\left( 
\begin{array}{cc}
\left( \ln q^{\ast }\right) _{x} & q \\ 
\kappa q^{\ast } & \left( \ln q\right) _{x}%
\end{array}%
\right) .
\end{equation}%
From the definition of $S$ and its decomposition the components of $\mathbf{s%
}$ are computed to 
\begin{equation}
s_{1}(x,t)=1+\frac{2\kappa \left\vert q\right\vert ^{4}}{q_{x}q_{x}^{\ast
}-\kappa \left\vert q\right\vert ^{4}},~~s_{2}(x,t)=\frac{\left\vert
q\right\vert ^{2}\left( q_{x}-\kappa q_{x}^{\ast }\right) }{q_{x}q_{x}^{\ast
}-\kappa \left\vert q\right\vert ^{4}},~~s_{3}(x,t)=i\frac{\left\vert
q\right\vert ^{2}\left( q_{x}+\kappa q_{x}^{\ast }\right) }{q_{x}q_{x}^{\ast
}-\kappa \left\vert q\right\vert ^{4}}.
\end{equation}%
It is trivial to verify that $\mathbf{s\cdot s}=1$. Thus for any solution to
the Hirota equation, with the additional constraint as specified in (\ref{co}%
), the vector $\mathbf{s}$ constitutes a solution to the ELL equation as
given in (\ref{ELL}).

One such solution we may employ is for instance the local one-soliton
solution obtained in \cite{CenFringHir}%
\begin{equation}
q(x,t)=\frac{(\mu +\text{$\mu $}^{\ast })^{2}\exp [\gamma +\mu x+\mu
^{2}t(i\alpha -\beta \mu )]}{(\mu +\text{$\mu $}^{\ast })^{2}+\exp [\gamma +%
\text{$\gamma $}^{\ast }+i\alpha t\left( \mu ^{2}-\text{$\mu $}^{\ast
2}\right) -\beta t\left( \mu ^{3}+\text{$\mu $}^{\ast 3}\right) +x(\mu +%
\text{$\mu $}^{\ast })]}.  \label{ones}
\end{equation}%
We briefly discuss some of the key characteristic behaviours of $\mathbf{s}$
for various choices of the parameters. When $\beta =0$, the solutions
correspond to the one-soliton solutions of the nonlinear Schr\"{o}dinger
equation. For real parameters $\mu $ we obtain the well known periodic
solutions to the ELL equation as seen in the left panel of figure \ref{Fig1}%
. However, when the shift parameters $\mu $ is taken to be complex we obtain
decaying solutions tending towards a fixed point.

\FIGURE{ \epsfig{file=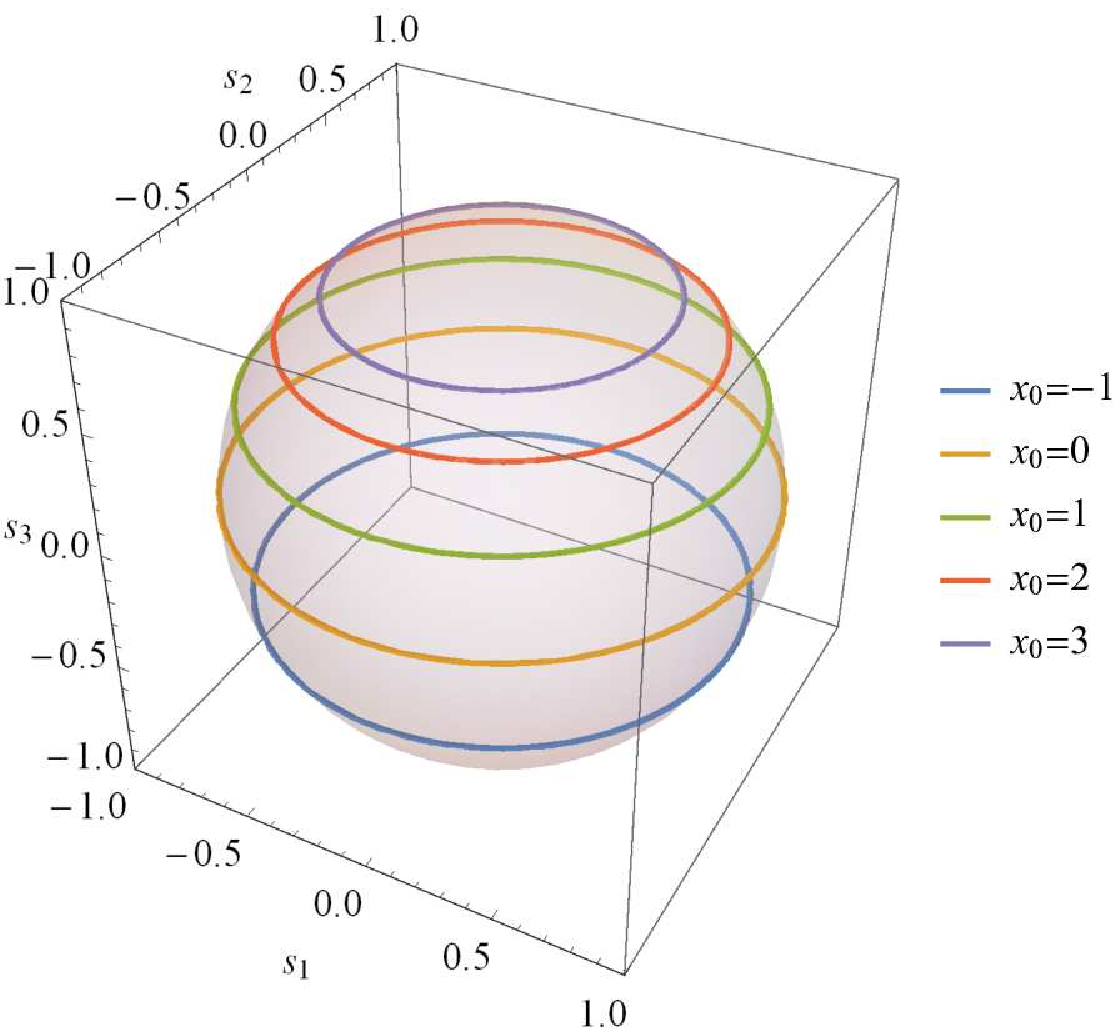, width=7.cm} \epsfig{file=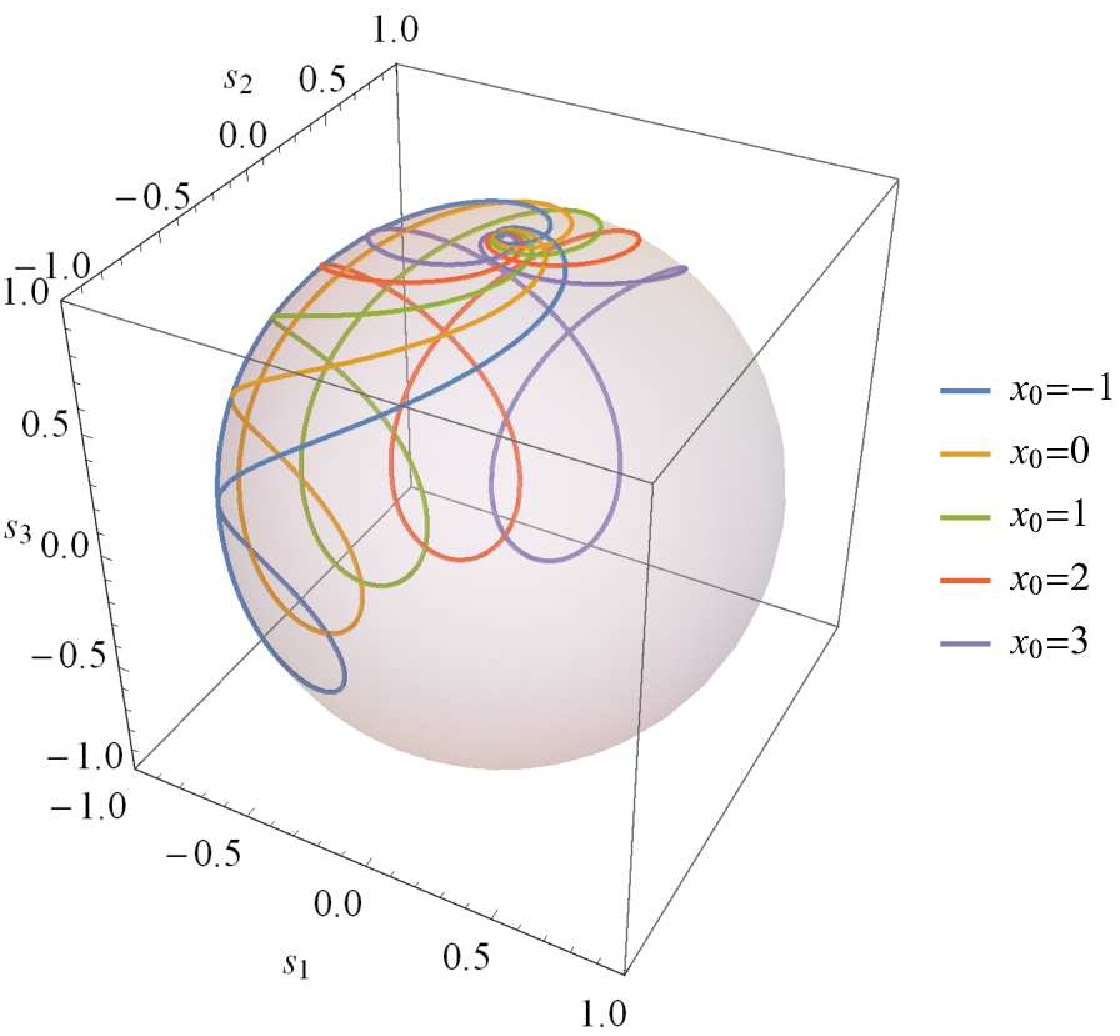, width=7.cm}
\caption{Local solutions to the extended Landau Lifschitz equation (\ref{ELL}) from a gauge
equivalent one-soliton solution (\ref{ones}) of the nonlinear Schr\"{o}dinger
equation for different initial values $x_{0}$, complex shift $\gamma =0.4+i0.2$, $\alpha =0.3$ and $\beta =0$. In the left panel the spectral parameter is real $\mu =0.3$ and
in the right panel it is complex $\mu =0.3+i0.1$.}
        \label{Fig1}}

When taking $\beta \neq 0$, that is the solutions to the Hirota equation
even for real values $\mu$ the bahaviour of the trajectories is drastically
changed even for small values of $\beta $, as they become more knotty and
convoluted as seen in the left panel of figure \ref{Fig2}. Complex values of 
$\mu$ are once more decaying solutions tending towards a fixed point.

\FIGURE{ \epsfig{file=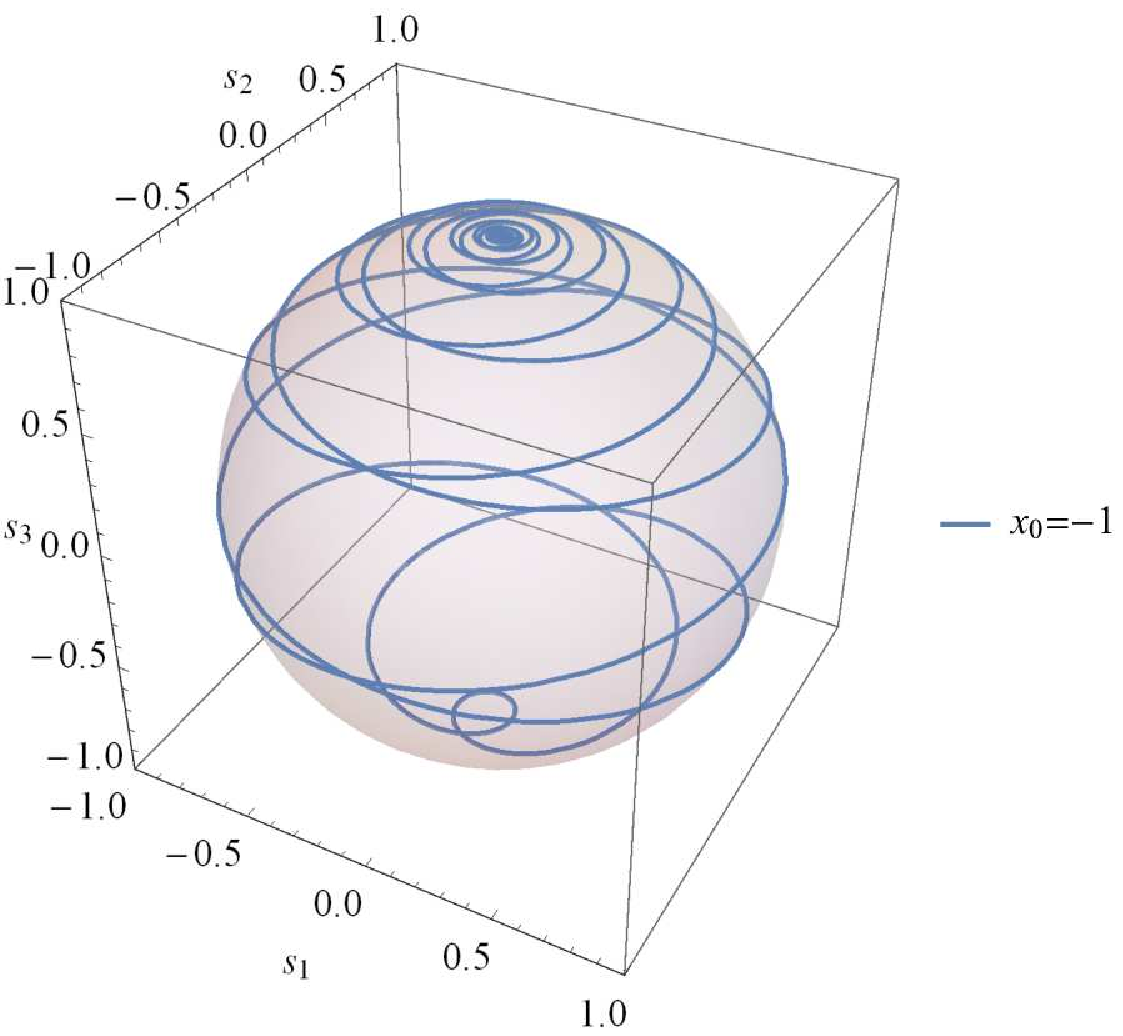, width=7.cm} \epsfig{file=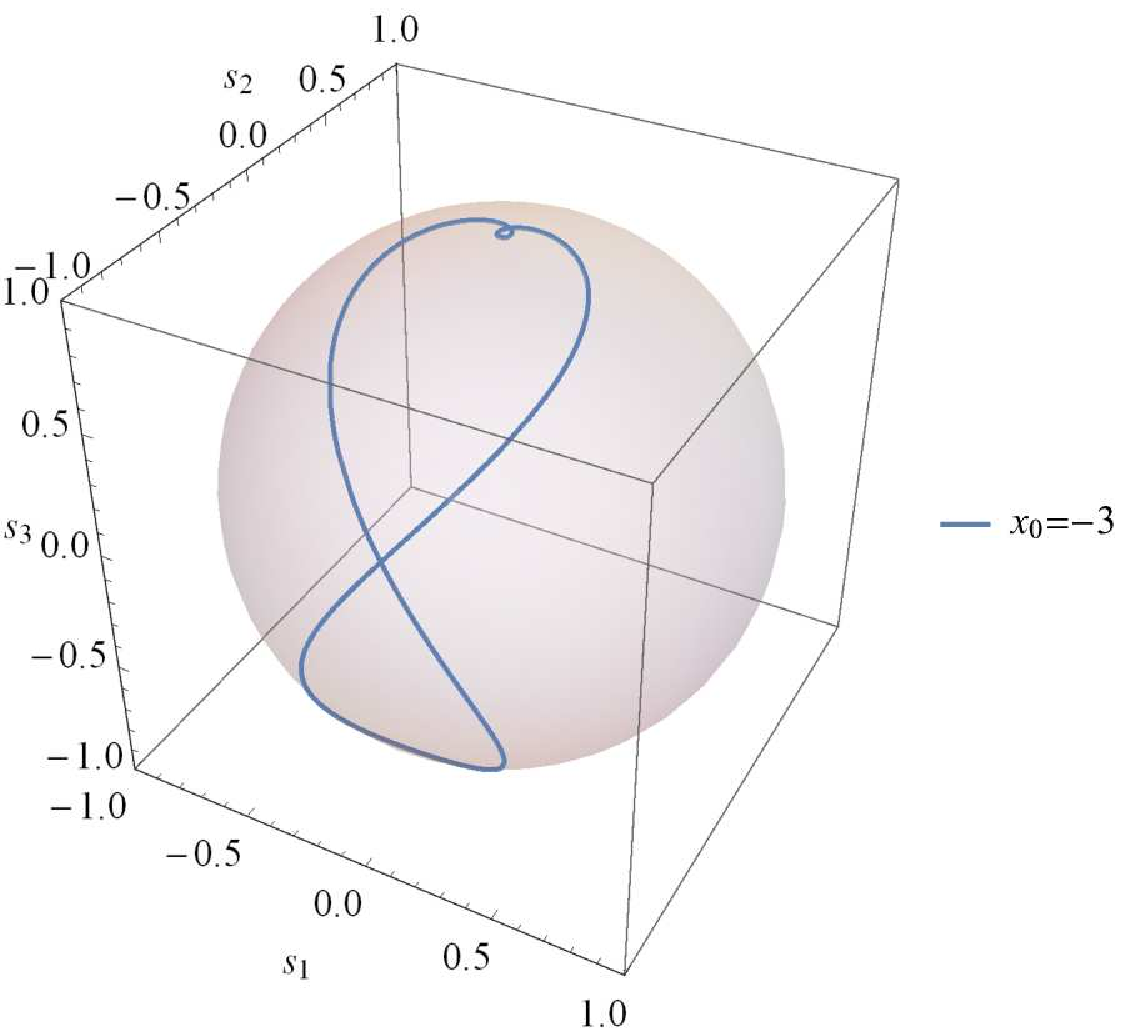, width=7.cm}
\caption{Local solutions to the extended Landau Lifschitz equation (\ref{ELL}) from a gauge equivalent
one-soliton solution (\ref{ones}) of the Hirota for a fixed value of $x_{0}$, complex shift $\gamma =0.4+i0.2$, $\alpha =0.3$ and $\beta =0.1$. In
the left panel the spectral parameter is real $\mu =0.3$ and in the right
panel it is complex $\mu =0.3+i0.1$.}
        \label{Fig2}}

Next we discuss the nonlocal solutions obtained by solving (\ref{11}), by
making the same additional assumption as for the construction of the local
solutions $b=cq(x,t)$ for some real constant $c$. In this case equation (\ref%
{nl1}) becomes%
\begin{equation}
a_{x}=-\kappa cq\tilde{q}^{\ast },\qquad a=c\left( \ln \tilde{q}^{\ast
}\right) _{x}.  \label{GG}
\end{equation}%
Now the compatibility between the two equations in (\ref{GG}) implies that%
\begin{equation}
\kappa q\tilde{q}^{\ast }=-\left( \ln \tilde{q}^{\ast }\right) _{xx}\quad
\Leftrightarrow \quad D_{x}^{2}f\cdot f=\kappa gh\text{,~~}\left( \ln
h\right) _{xx}=0.\text{ }  \label{con}
\end{equation}%
with $q=g/f$ , $f,g\in \mathbb{C}$ and $h=2f~\tilde{g}^{\ast }/\tilde{f}%
^{\ast }$. Once again the first relation on the right hand side in (\ref{con}%
) occurs in the bilinearisation of the nonlocal Hirota equation, see section
4.1 \cite{CenFringHir}. However, as for the local case in (\ref{co}) the
second relation is an additional constraint that is not automatically
satisfied by all solutions. We have therefore obtained a solution for the
nonlocal gauge field operator in the form%
\begin{equation}
G=c\left( 
\begin{array}{cc}
\left( \ln \tilde{q}^{\ast }\right) _{x} & q \\ 
\tilde{q}^{\ast } & \left( \ln q\right) _{x}%
\end{array}%
\right) ,
\end{equation}%
so that the matrix $S$ can be computed directly from its defining relation (%
\ref{SG}). Using the expansion for $S$ in terms of the components of $%
\mathbf{s}$ we compute 
\begin{equation}
s_{1}(x,t)=\frac{q^{2}\tilde{q}_{x}^{\ast }-\tilde{q}^{\ast 2}q_{x}}{\tilde{q%
}^{\ast 2}q^{2}-\tilde{q}_{x}^{\ast }q_{x}},~~s_{2}(x,t)=i\frac{q^{2}\tilde{q%
}_{x}^{\ast }+\tilde{q}^{\ast 2}q_{x}}{\tilde{q}^{\ast 2}q^{2}-\tilde{q}%
_{x}^{\ast }q_{x}},~~s_{3}(x,t)=-\frac{q^{2}\tilde{q}^{\ast 2}+q_{x}\tilde{q}%
_{x}^{\ast }}{\tilde{q}^{\ast 2}q^{2}-\tilde{q}_{x}^{\ast }q_{x}}.  \label{s}
\end{equation}%
It is trivial to verify that $\mathbf{s\cdot s}=1$. Thus for any solution to
the nonlocal Hirota equation, with the additional constraint as specified in
(\ref{con}), the complex valued vector $\mathbf{s}$ constitutes a solution
to the ELLE as given in (\ref{ELL}).

One such solution one may employ is the nonlocal one-soliton solution
obtained in \cite{CenFringHir}%
\begin{equation}
q(x,t)=\frac{(\mu -\mu ^{\ast })^{2}\exp \left[ \gamma +\mu (x+i\mu t(\alpha
-\delta \mu ))\right] }{(\mu -\mu ^{\ast })^{2}+\exp \left[ \gamma +\text{$%
\gamma $}^{\ast }+it\left( \alpha (\mu ^{2}-\mu ^{\ast 2})+\delta \left( \mu
^{\ast 3}-\mu ^{3}\right) \right) +x(\mu -\mu ^{\ast })\right] }.
\label{onenl}
\end{equation}%
Let us analyze how $\mathbf{m}$ and $\mathbf{l}$ behave in this case. As
expected, the trajectories will no stay on the unit sphere. However, for
certain choices of the parameters it is possible to obtain well localized
closed three dimensional trajectories that trace out curves with fixed
points at $t=\pm \infty $ as seen for an example in figure \ref{Fig3}. Thus
the nonlocal nature of the solutions to the Hirota equation has apparently
disappeared in the setting of the extended Landau Lifschitz equation.
However, not all solutions are of this type as some of them are now
unbounded.

\FIGURE{ \epsfig{file=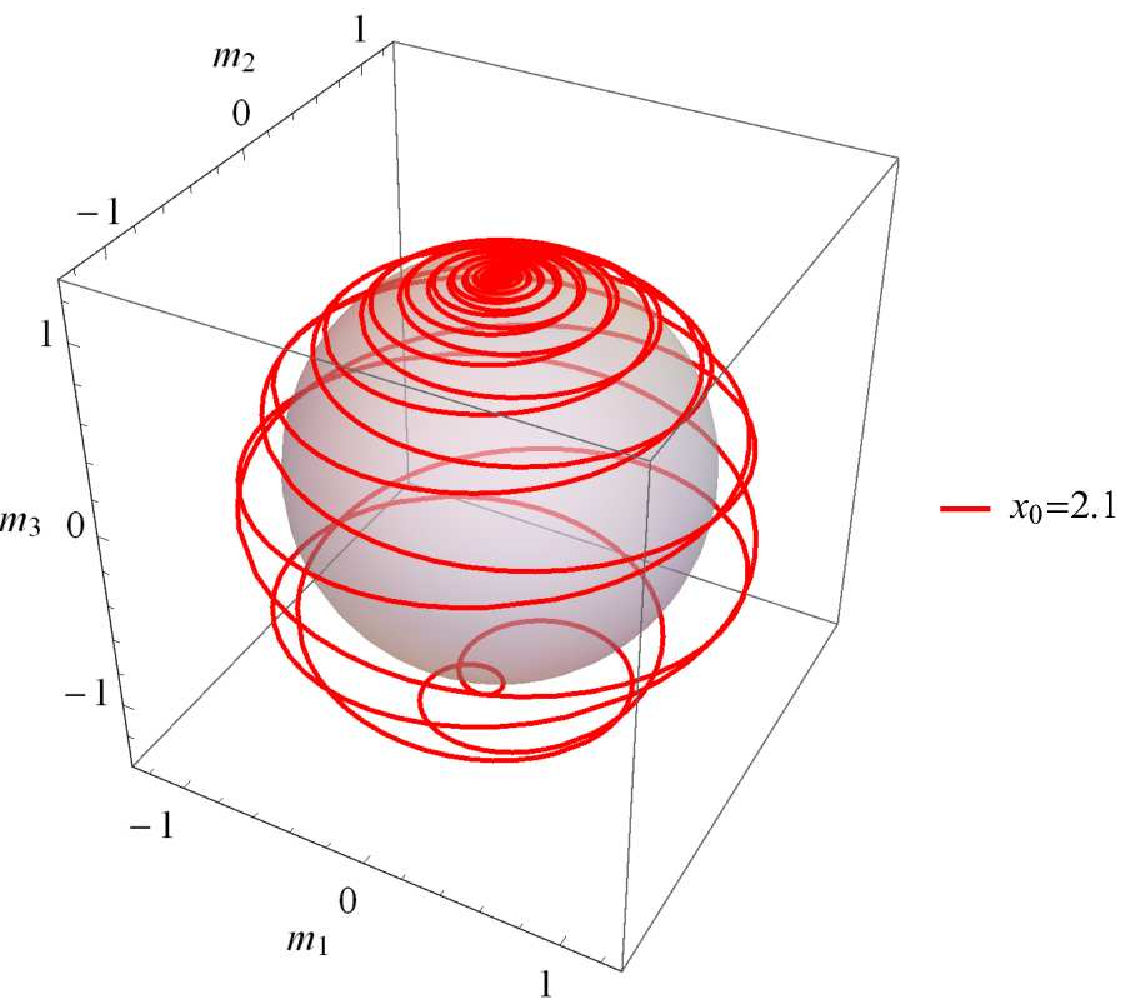, width=7.cm} \epsfig{file=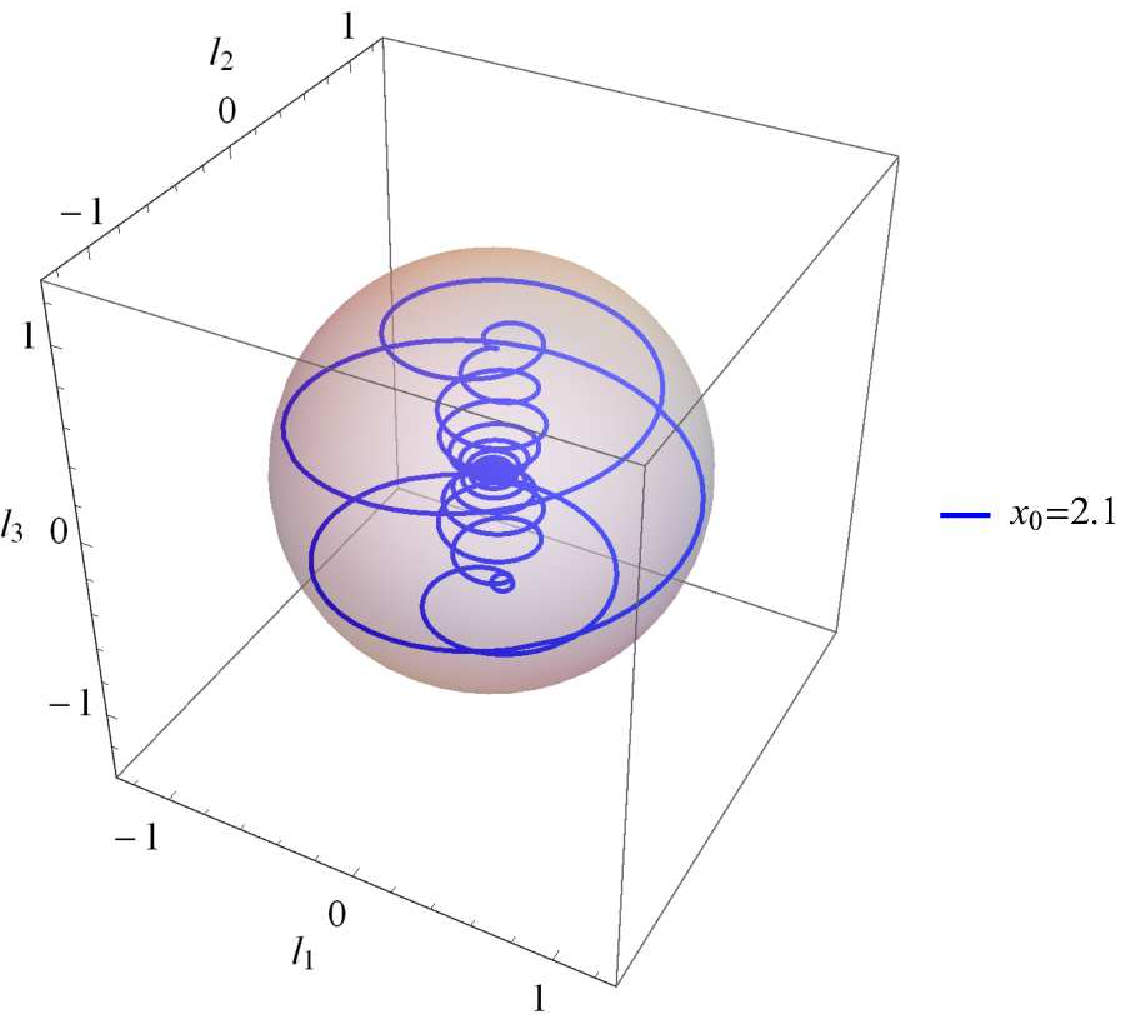, width=7.cm}
\caption{Nonlocal solutions to the extended Landau Lifschitz equation (\ref{NELL1}) and (\ref{NELL2}) from a gauge equivalent
one-soliton solution (\ref{onenl}) of the Hirota equation for a fixed value of
$x_{0}$, vanishing complex shift $\gamma =0$, $\mu = i 0.55$, $\alpha =1.5$ and $\delta =0.15$.}
        \label{Fig3}}

\subsection{Nonlocal two-soliton solutions to the ELLE}

While the computation of the solutions to the ELL equation is
straightforward when computing $G$ directly with some additional
constraints, not all Hirota solutions obey them. Let us therefore use the
two-soliton solution (\ref{2s}) in the representation (\ref{m1})-(\ref{l3})
to study the nonlocal two-soliton solutions to the ELL equation. The
two-soliton structure is best revealed when plotting it for fixed time over
space. In figure \ref{Figtwos} we show each component of $\mathbf{m}$ and $%
\mathbf{l}$ separately, displaying clearly two distinct one-soliton
structures.

\FIGURE{ \epsfig{file=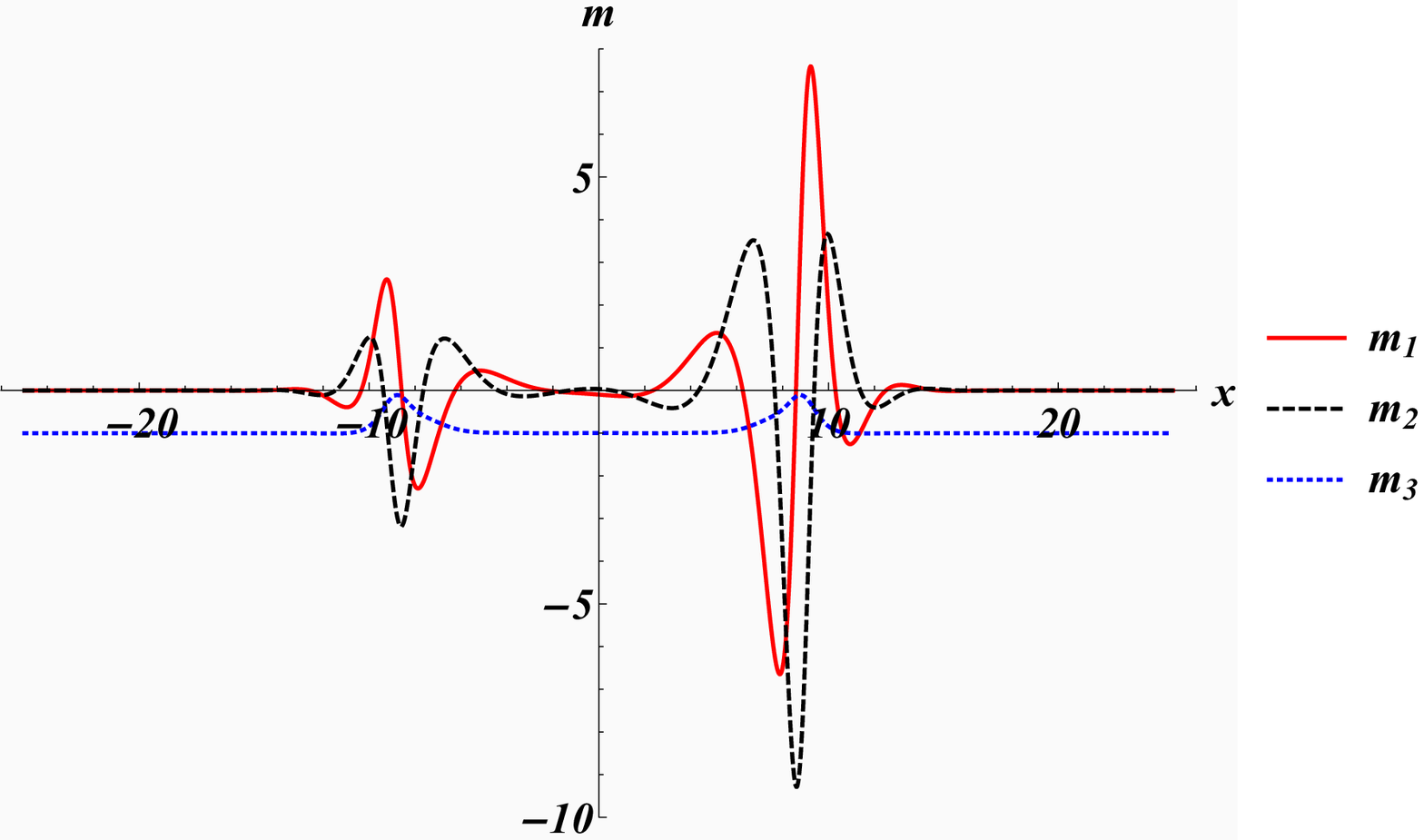, width=7.0cm} \epsfig{file=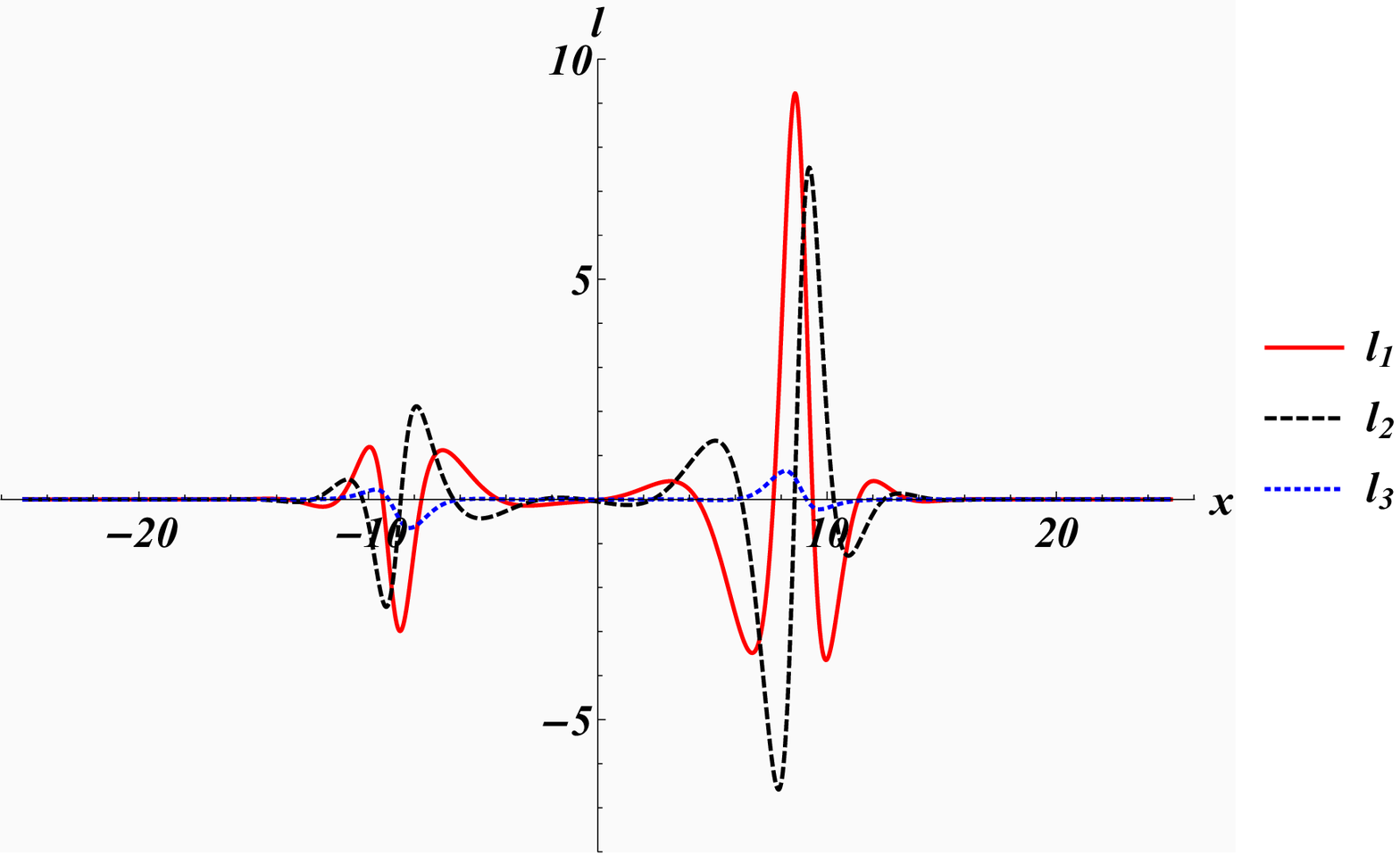, width=7.cm} 
\caption{Nonlocal two-soliton solutions to the extended Landau Lifschitz equation for fixed time $t=3$ as a
function of with parameters $\alpha =1.2$, $\delta =0.2$, $\kappa =3$, $\lambda =0.4-i0.3$, $\rho =0.7+i0.5$, $\gamma _{1}=i5.1$, $\gamma _{2}=i0.1$, $\gamma _{3}=-i1.1$ and $\gamma _{4}=i0.2$.}
        \label{Figtwos}}

\section{Conclusions}

We discussed two different types of local/nonlocal gauge transformations:
The first of them, $G,$ relates the auxiliary functions in the two spectral
problems of the local/nonlocal extended continuous limit of the Heisenberg
equation to the local/nonlocal Hirota equation. The explicit form of the
gauge functions can be used to establish a concrete relation between
solutions of one system to the other. This concrete map when applied to
solutions works most efficiently in one direction from the spectral problem
the local/nonlocal extended continuous Heisenberg equation to solutions of
the (nonlocal) Hirota equation, as stated explicitly in (\ref{qn}) and (\ref%
{rn}). This map is not easily invertible and instead we used (\ref{co}) and (%
\ref{con}) to provide an alternative. While we demonstrated that the (\ref%
{co}) and (\ref{con}) are equivalent to equations that emerge in the
bilinearization process, they also require an additional constraint that is
not satisfied by all solutions, so that not all solutions are obtainable in
this manner. The second type of gauge transformation, $\hat{G}$, is an
auto-gauge transformation that relates the auxiliary functions in the
spectral problem of the (nonlocal) extended continuous Heisenberg equation
to itself. This gauge transformation can be interpreted as a Darboux
transformation and allows to construct a new solution from a known one. In
an analogous fashion to Darboux-Crum transformations, it can be iterated to
produce multi-soliton solutions.

The nonlocality can be implemented separately in the two systems by applying
a parity complex conjugation map to different sets of equations. For the
Hirota system it is most naturally applied to the pair of equations (\ref%
{zero1}) and (\ref{zero2}), resulting from the zero curvature formulation as
discussed in \cite{CenFringHir}. For the extended version of continuous
limit of the Heisenberg equation it is most obviously applied to its
component version (\ref{uv1}) and (\ref{uv2}). The two versions of
nonlocality in the two systems were shown to be related to each by means of
the gauge transformation $G$ as demonstrated by (\ref{nonloc}). As
demonstrated, one may, however, also map components of the gauge
transformation matrix to each other in a consistent manner. At the level of
the spectral problem the nonlocality is implemented via mapping components
of the seed functions consistently to each other.

Various issues are worthy of further exploration. It is well known \cite%
{hasimoto1921,lamb1976solitons,lakshmanan1,lakshmanan2} that the Landau
Lifschitz equation, i.e. (\ref{ELL}) for $\beta =0$, admits a geometric
interpretation that directly relates the curvature and torsion of a vector
field to any solution of the nonlinear Schr\"{o}dinger equation when
expressed in form of the Hasimoto map \cite{hasimoto1921}. In \cite%
{JFAinprep} we demonstrate that these relations and interpretations can be
extended to the extended nonlocal versions of this equation. Naturally it
would also be interesting to explore the behaviour of the systems arising
from the other types of $\mathcal{PT}$-conjugation as discussed in \cite%
{CenFringHir}.

\medskip

\noindent \textbf{Acknowledgments:} JC is supported by a City, University of
London Research Fellowship. FC was partially supported by Fondecyt grant
1171475 and would like to thank the Department of Mathematics at City,
University of London and the Departamento de F\'isica Te\'orica, \'Atomica y
\'Optica at the Universidad de Valladolid for kind hospitality. AF would
like to thank the Instituto de Ciencias F{\'{\i}}sicas y Matem{\'{a}}ticas
at the Universidad Austral de Chile for kind hospitality and Fondecyt for
financial support.

\newif\ifabfull\abfulltrue


\medskip

\end{document}